\documentclass{emulateapj}

\usepackage{natbib} 
\begin{document}
\def\mpch {$h^{-1}$ Mpc} 
\def\kpch {$h^{-1}$ kpc} 
\def\kms {km s$^{-1}$} 
\def\lcdm {$\Lambda$CDM } 
\def\xir {$\xi(r)$}
\def\wprp {$w_p(r_p)$}
\def\xisp {$\xi(r_p,\pi)$}
\def\xis {$\xi(s)$}
\def\rr {$r_0$}
\def\etal {et al.}

\title{AEGIS: The Clustering of X-ray AGN Relative to Galaxies at $z\sim1$}
\author{Alison L. Coil\altaffilmark{1,2,3}, 
Antonis Georgakakis\altaffilmark{4,5},
Jeffrey A. Newman\altaffilmark{6},
Michael C. Cooper\altaffilmark{2}, 
Darren Croton\altaffilmark{7}, 
Marc Davis\altaffilmark{8},
David C. Koo\altaffilmark{9},
E.S. Laird\altaffilmark{4},
K. Nandra\altaffilmark{4},
Benjamin J. Weiner\altaffilmark{2},
Christopher N. A. Willmer\altaffilmark{2}, 
Renbin Yan\altaffilmark{10}, 
}
\altaffiltext{1}{Hubble Fellow}
\altaffiltext{2}{Steward Observatory, University of Arizona,
Tucson, AZ 85721}
\altaffiltext{3}{Department of Physics, University of California,
San Diego, CA 92093}
\altaffiltext{4}{Astrophysics Group, Blackett Laboratory, Imperial College, London SW7 2BZ, UK}
\altaffiltext{5}{National Observatory of Athens, V. Paulou \& I. Metaxa, 11532, Greece}
\altaffiltext{6}{Department of Physics and Astronomy, University of Pittsburgh, Pittsburgh, PA 15260}
\altaffiltext{7}{Centre for Astrophysics \& Supercomputing, Swinburne University of Technology, P.O. Box 218, Hawthorn, VIC 3122, Australia}
\altaffiltext{8}{Department of Astronomy, University of California, Berkeley, CA 94720}
\altaffiltext{9}{University of California Observatories/Lick
Observatory, Department of Astronomy and Astrophysics, University of
California, Santa Cruz, CA 95064}
\altaffiltext{10}{Department of Astronomy and Astrophysics, University of Toronto, Toronto, ON  M5S 3H4, Canada}

\begin{abstract}

We measure the clustering of non-quasar 
X-ray AGN at $z=0.7-1.4$ in the AEGIS field.  
Using the cross-correlation of 113 {\it Chandra}-selected AGN, 
with a median log $L_{\rm X}=42.8$ erg $s^{-1}$, with 
$\sim$5,000 DEEP2 galaxies, we find that the X-ray AGN are fit by a 
power law with a clustering scale length of 
$r_0=5.95 \pm0.90$ \mpch \ and slope $\gamma=1.66 \pm0.22$.  
X-ray AGN have a similar clustering amplitude as red, quiescent and `green' transition
galaxies at $z\sim1$ and are significantly more clustered than blue, star-forming galaxies.  
The X-ray AGN clustering strength is primarily determined by the host galaxy
color; AGN in red host galaxies
are significantly more clustered than AGN in blue host galaxies, with a relative bias
that is similar to that of red to blue DEEP2 galaxies.  We detect no dependence
of clustering on optical brightness, X-ray luminosity, or hardness ratio  
within the ranges probed here.  We find evidence for galaxies hosting
 X-ray AGN to be more clustered than a sample of galaxies with matching joint 
optical color and magnitude distributions.  This implies that galaxies hosting
X-ray AGN are more likely to reside in groups and more massive dark matter 
halos than galaxies of the same color and luminosity without an 
X-ray AGN.  
In comparison to 
optically-selected quasars in the DEEP2 fields, we find that X-ray AGN 
at $z\sim1$ are more clustered than optically-selected quasars (with a 2.6$\sigma$ 
significance) and therefore may reside in more massive dark matter halos.
 Our results are consistent with galaxies undergoing a quasar
phase while in the blue cloud before settling on the red sequence with a 
lower-luminosity X-ray AGN, {\it if} they are similar objects at different
evolutionary stages.

\end{abstract}


%_______________________________________________________________________

\section{Introduction}

% motivation

It has recently become clear that black holes play a key role
in the evolution of galaxies. It is now
understood that most galaxies have supermassive black holes in their
nuclei \citep[for reviews, see e.g.][]{Richstone98,Ferrarese05} and that these
black holes power active galactic nuclei (AGN).  There are strong
observed correlations between black hole mass and galaxy properties
such as the stellar velocity dispersion in the bulge
\citep{Gebhardt00,Ferrarese00}, that indicate some form of feedback
or connection between the growth of black holes and their parent
galaxies. 

It remains unclear how galaxies and AGN co-evolve and what impact the
AGN have on the evolution of their host galaxies.  It is also not known
 what the accretion mechanism is for AGN and what their fueling source is.
Measuring the environments and clustering properties of AGN is key to 
understanding their nature: constraining how they are triggered and fueled, 
inferring their lifetimes, and
determining which galaxy populations host different kinds of AGN (quasars,
Seyferts, radio-loud AGN, etc.).
Additionally, clustering measurements constrain the masses of the dark matter
halos which host AGN and allow various types of AGN to be placed in a 
cosmological context.
These pieces are all necessary to understand the relevance of AGN 
for galaxy evolution.

Cosmological numerical simulations, both hydrodynamic and
semi-analytic, predict the co-evolution of black holes
and galaxies as well as the lifetimes and fueling mechanisms for AGN.
In these simulations the AGN physics itself is not well resolved and 
often assumptions are made regarding the details of accretion 
and feedback from the AGN.
Most models of high luminosity AGN assume that mergers drive the AGN
accretion \citep[e.g.,][]{Kauffmann00, Springel05a}, though 
lower-luminosity AGN may be fueled in other ways \citep{Hopkins06}.
Recent theoretical models often include some form of AGN feedback,
motivated by the observational links between the properties of galaxies and their
central black holes; this feedback can cause a decline in the star
formation activity of massive galaxies, creating red ellipticals.  
Many models find that AGN
feedback of some sort is in fact required to create the red sequence of quiescent 
galaxies (though see \cite{Naab07} and \cite{Dekel08} for counter
examples). For massive galaxies, accretion of gas from cooling
flows in dense environments may produce relatively low-luminosity AGN that 
in turn heat the bulk of the cooling gas and prevent it from
falling into the galaxy's center to form stars 
\citep[e.g.,][]{Cattaneo06,Croton06}.  Other models propose that 
mergers trigger quasars that
drive outflows that sweep away the gas, not allowing it to form stars
\citep[e.g.,][]{Springel05a}. 

These theoretical models can be tested
using observations of the location of AGN within the web of 
large-scale structure.  Measuring the clustering of both low and high luminosity AGN is
crucial for determining if their fueling mechanisms differ and searching for links 
between AGN fueling and local environment.  To measure clustering
properties accurately requires large samples to overcome both Poisson errors
and cosmic variance. 

It is now known that optical surveys fail to detect a large fraction of 
the full AGN population, particularly weak or obscured AGN 
\citep[e.g.,][]{Mushotzky00,Alexander01,Mushotzky04,Brandt05}.  
X-ray detection is generally recognized as a more robust way to 
obtain a relatively unbiased AGN sample, though it does not detect all AGN 
\citep[e.g.,][]{Donley05, Heckman05}, especially optically-varying AGN
\citep{Morokuma08}. As soft X-ray samples can miss 
obscured AGN \citep[e.g.,][]{Comastri04}, an ideal sample includes objects
detected in either the hard or soft bands; however, even the deepest of 
the current X-ray surveys are likely to miss the most heavily obscured 
``Compton thick'' AGN \citep{Gilli07}.

Most previous attempts to measure the clustering of X-ray AGN in 
{\it ROSAT, XMM-Newton} and {\it Chandra} surveys have been fundamentally 
limited by a lack of spectroscopic redshifts, resulting in conflicting 
values for the inferred 
correlation amplitude.  Most studies measure the projected
angular clustering of X-ray sources 
\citep{Akylas00,Yang03,Basilakos04,Miyaji07,Plionis08}; to interpret the
angular clustering the redshift distribution of the sources, $dN/dz$, must be known
to fairly high precision, as the inferred correlation length is particularly
sensitive to the range of redshifts spanned.  Lacking a large
sample of spectroscopic redshifts for these sources, most authors assume 
a redshift distribution using the observed X-ray luminosity function and 
AGN population synthesis models.  However, errors in the $dN/dz$ distribution
are generally not propagated to the inferred correlation length, so that the quoted
errors on the clustering amplitude are highly underestimated and often
the inferred correlation lengths are extremely large 
(e.g., $r_0=9-19$ \mpch \ in \cite{Basilakos05}). 
Additionally, most studies do not include cosmic variance errors, which can 
be large for the relatively small fields with deep X-ray data.

\begin{figure*}[t]
\epsscale{1.0}
\plotone{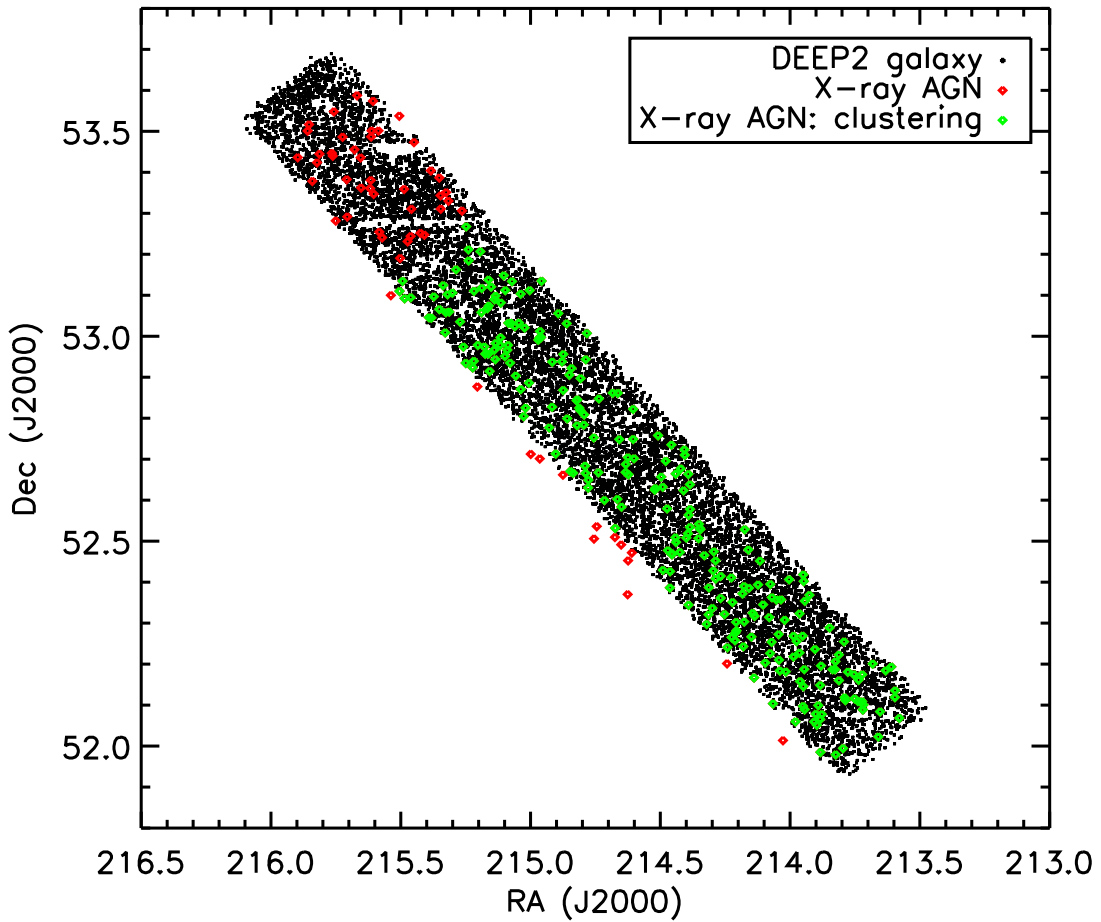}
\caption{Spatial distribution of DEEP2 galaxies (black squares) and X-ray 
AGN (colored diamonds) with $0.2<z<1.4$ in the Extended Groth Strip field.  
X-ray AGN used for the clustering analysis
performed here are shown in green; these sources are within the boundary of
the DEEP2 survey in this field and cover the lower three quadrants of the 
strip, where the DEEP2 selection function is uniform 
(see Section 3.2 for details).}
\label{skydist}
\end{figure*}

Recently, several studies have employed spectroscopic redshifts to measure 
the direct correlation function of X-ray AGN.  \cite{Mullis04} use data 
from the {\it ROSAT} North Ecliptic Pole survey to measure the clustering
of 219 bright ($L_{\rm x}=9.2 \times 10^{43} \ h_{70}^{-2} erg \ s^{-1}$), 
soft (0.5-2.0 keV) X-ray sources 
at $z\sim0.2$ on large scales, $r=5-100$ \mpch, finding a correlation 
length of $r_0=7.4 \pm1.9$ \mpch.  \cite{Gilli05} measure the clustering
of X-ray AGN at $z=0-4$ in both of the $\sim$0.1 sq. degree 
{\it Chandra} Deep Fields (CDF), 
finding $r_0=10.3 \pm1.7$ \mpch \ and $\gamma=1.33 \pm0.14$ for 240 sources 
in the northern field and $r_0=5.5 \pm0.6$ \mpch \ and $\gamma=1.50 \pm0.12$ 
for 124 sources in the southern field.  These error bars do 
not include cosmic variance, however, which is severe for the small 
CDF fields and most likely accounts for the discrepancy between the measurements.  
\cite{Yang06} measure the clustering of 233 spectroscopic sources at $z=0.1-3$ 
in the 0.4 sq. degree {\it Chandra} CLASXS area, as well as 252 sources 
in the CDFN field.  They find $r_0=5.7 +0.8/-1.5$ \mpch \ in CLASXS and 
$r_0=4.1 +0.7/-1.1$ \mpch \ in the CDFN field, where the errors are from
bootstrap resampling of the data. More recently, \cite{Gilli09} find 
in the 2 sq. degree 
XMM-COSMOS field that 349 X-ray AGN with $0.4<z<1.6$ have a clustering
scale length of $r_0=5.2 \pm1.0$ \mpch \ and $\gamma=1.7 \pm0.2$, where the
error bars quoted include only Poisson errors.

Ideally, one would like to measure the correlation amplitude of AGN in 
a relatively narrow redshift range using spectroscopic redshifts, 
and compare with the clustering of galaxies 
in the same volume to understand the relationship of AGN to their 
host galaxies and avoid much of the cosmic variance error.  
In this paper, we measure the clustering properties of 113 X-ray
selected AGN in the redshift range $0.7<z<1.4$ (a subset of the 463 X-ray 
objects in this field with spectroscopic redshifts) using deep {\it Chandra} data
in the Extended Groth Strip (EGS).  
For comparison, the only other X-ray field that has a large sample of
spectroscopic redshifts and is comparable in both size and depth to
the EGS is the XMM-COSMOS field.  The environments of 53 X-ray selected
AGN in the EGS were measured in \cite{Georgakakis07} and compared to 
galaxies in the same volume; they found that X-ray AGN avoid under-dense
regions and have environments similar to those of optically red galaxies.
Here we measure the clustering of a larger sample of AGN in the same field, 
as a function of scale, and compare to galaxies in the same volume. 
To measure the clustering amplitude of the AGN we do not use 
the auto-correlation function of the AGN, as it is very sensitive to 
spatially-varying selection functions (which is troublesome given that the
{\it Chandra} point spread function and telescope effective area vary across
the field of view) and is subject to large shot noise due 
 to the relatively small AGN sample size.
Instead, we take advantage of the large, uniformly selected DEEP2 galaxy
survey in the EGS and measure the clustering of more abundant galaxies 
around the AGN using the two-point cross-correlation function and then divide by the
correlation function of the galaxies themselves to determine the
AGN clustering strength. 
We measure the clustering of the AGN on both small ($<1$ \mpch) and
large ($1-10$ \mpch) scales and compare the results to the clustering
of galaxies from the DEEP2 galaxy redshift survey 
as a function of color and luminosity at the same
redshift, to understand which galaxies host these AGN and to constrain
their host dark matter halo masses.  By comparing these clustering
measurements with those for optically-selected samples of luminous AGN, we 
can investigate the 
relationship between accretion activity and large-scale structure,
which strongly constrains theoretical models of the interplay between
AGN and galaxy evolution.

The outline of the paper is as follows: \S2 briefly describes the
{\it Chandra} and DEEP2 surveys in the EGS and presents additional
spectroscopic follow-up of X-ray sources taken with the MMT. In \S3 
we discuss the optical properties of the X-ray AGN and 
define the various AGN and galaxy samples used here. \S4 outlines the 
methods used to measure the cross-correlation function and infer the
real-space auto-correlation function.  Results for various AGN and 
galaxy samples are shown in \S5, as a function of optical and 
X-ray properties.  These results are discussed in \S6, and we conclude 
in \S7. 
To convert measured redshifts to comoving distances along the line of
sight, we assume a flat \lcdm cosmology with $\Omega_{\rm m}=0.3$ and
$\Omega_{\Lambda}=0.7$.  We define $h \equiv {\rm {\it H}_0/(100 \ km
\ s^{-1} \ Mpc^{-1}})$ and quote correlation lengths, \rr, in comoving
\mpch.

%_______________________________________________________________________

\section{Data}

\begin{figure*}[t]
\epsscale{1.1}
\plottwo{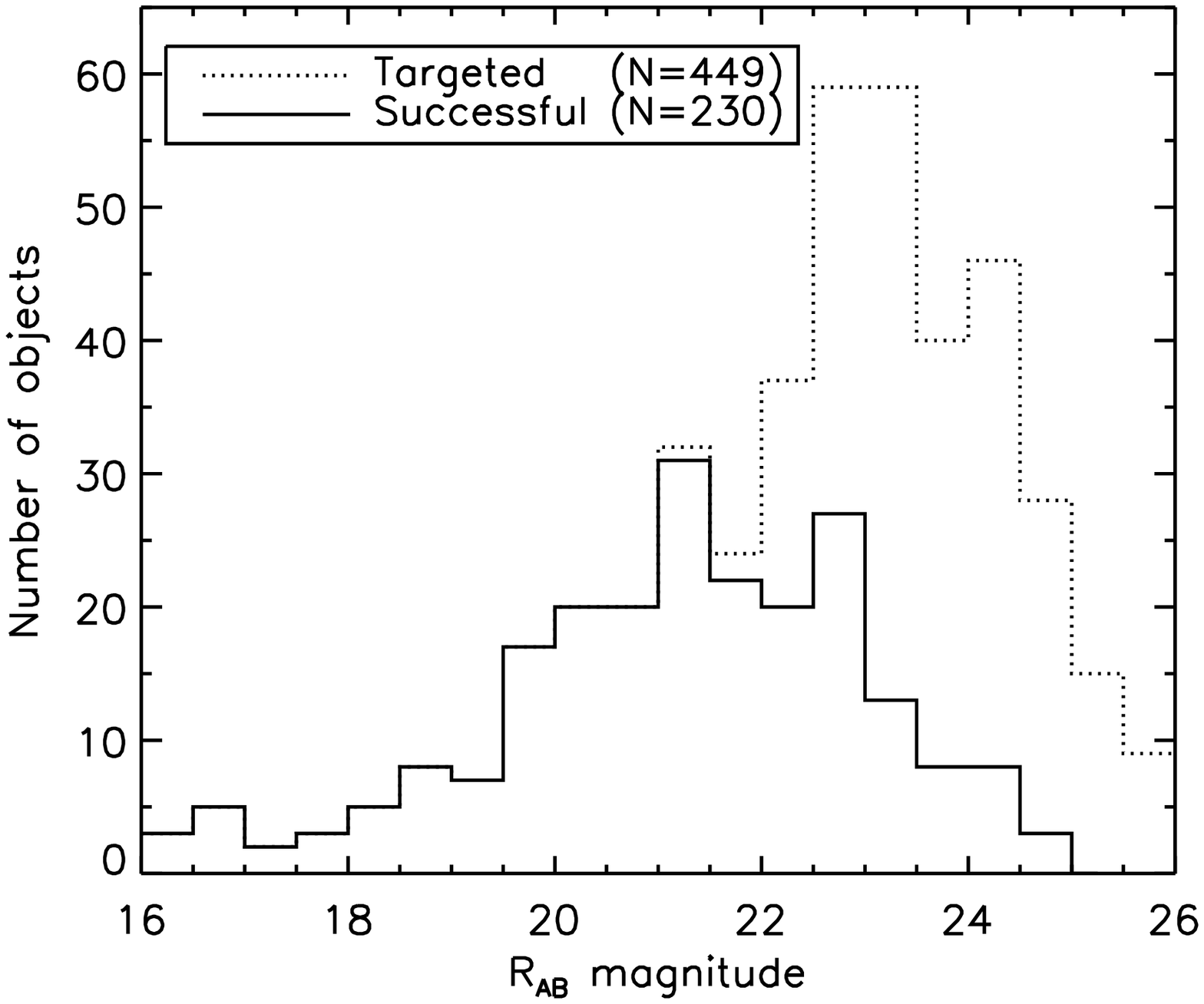}{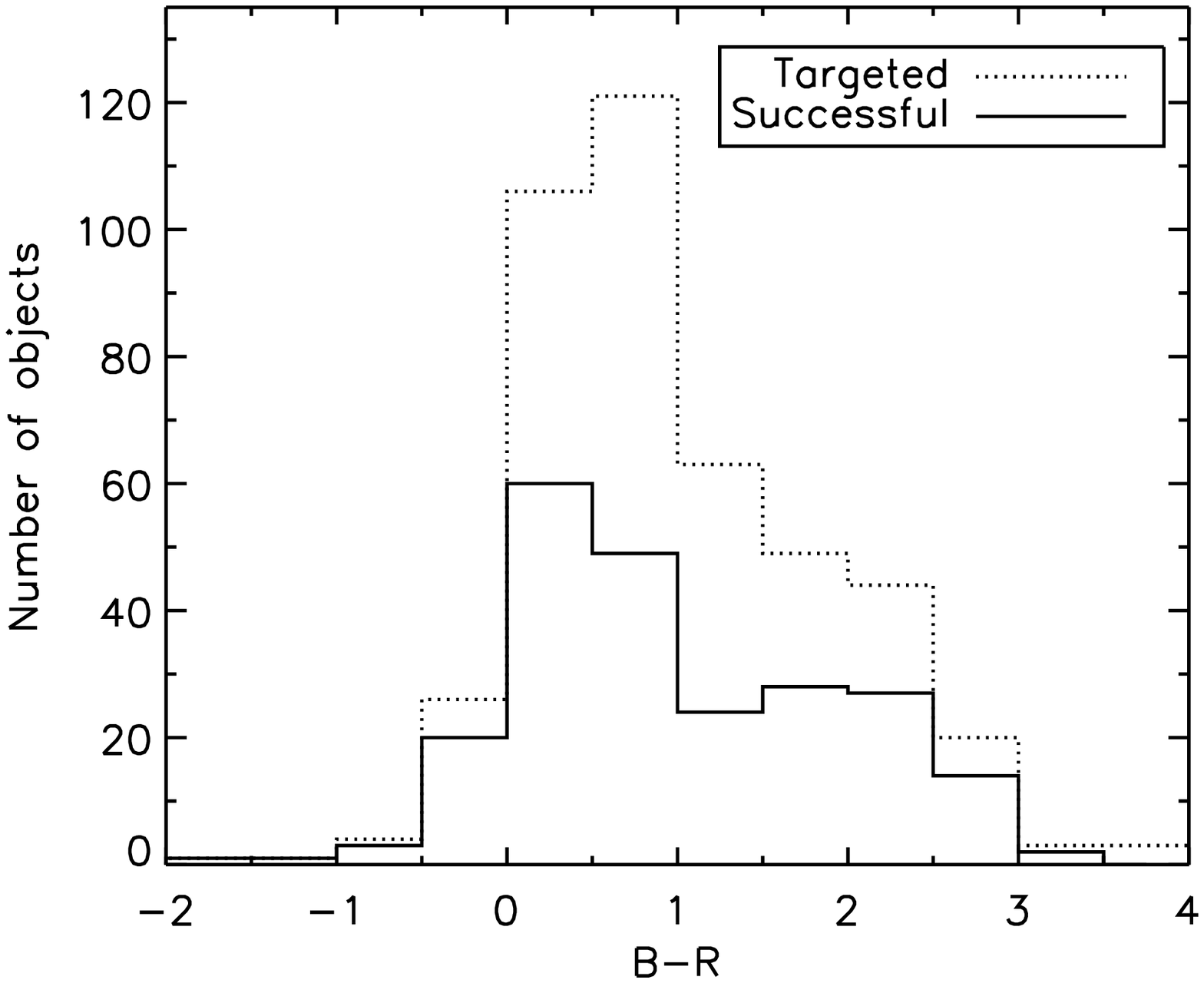}
\caption{Left: Optical $R_{\rm AB}$ magnitude distribution for X-ray sources targeted 
(dotted line) for MMT/Hectospec spectroscopic observations.  The solid
 line shows objects which yielded a high-confidence spectroscopic 
redshift.  Right: Optical apparent $B-R$ colors 
for X-ray sources targeted (dotted line) and for those with 
a high-confidence redshift (solid line).  
}
\label{mmt_mags}
\end{figure*}

\begin{figure*}[t]
\epsscale{1.1}
\plottwo{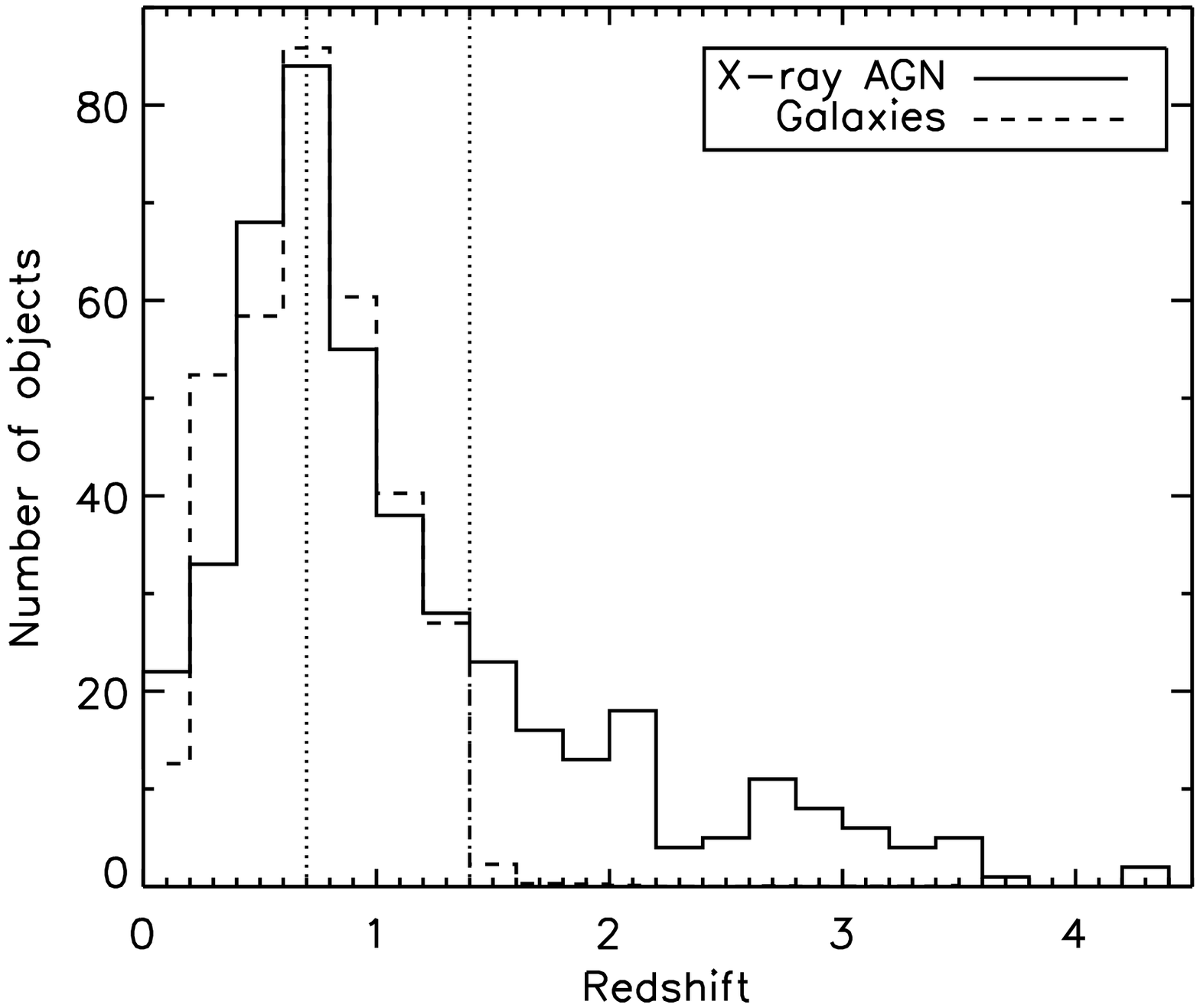}{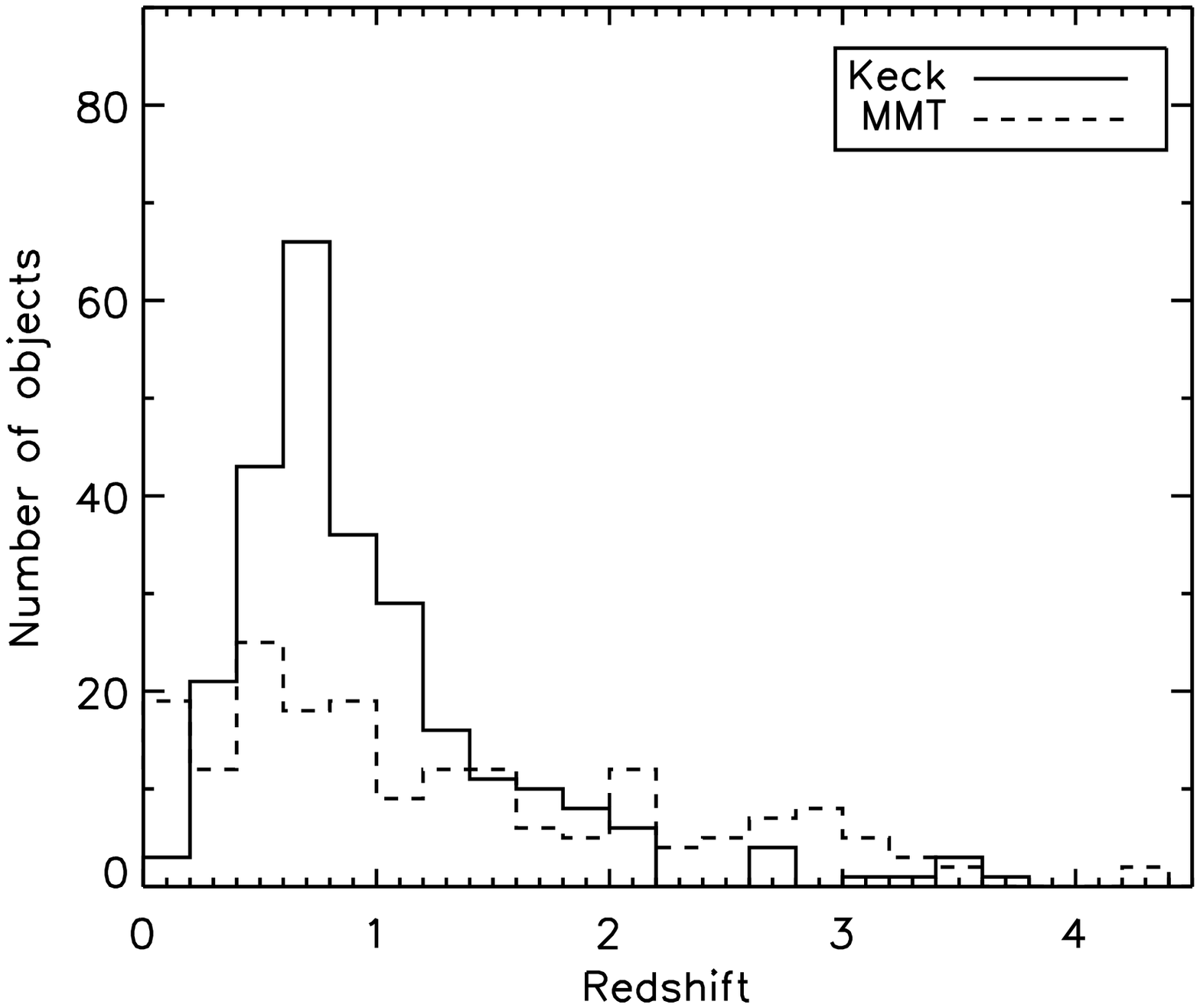}
\caption{Left: Redshift distribution of all 463 {\it Chandra} 
X-ray sources in AEGIS with spectroscopic
redshifts (solid line) and DEEP2 galaxies in the EGS (dashed line).  
The dotted vertical lines show the redshift range used here to  
measure the clustering properties of X-ray AGN and compare with galaxies.
Right: The redshift distribution of X-ray sources where the redshift was
measured in Keck/DEIMOS spectra (solid line, 233 sources) or 
MMT/Hectospec spectra (dashed line, 230 sources).}
\label{mmt_z}
\end{figure*}

\subsection{{\it Chandra} Imaging}

X-ray AGN are identified using deep {\it Chandra} data \citep{Laird09} 
obtained as part of 
the AEGIS survey of the Extended Groth Strip \citep[EGS,][]{Davis07}. 
This dataset consists of 8 Chandra ACIS-I pointings,
each with a total integration time of about 200\,ks split in at least
3 shorter exposures obtained at different epochs.  The data reduction,
source detection and flux estimation are described in detail by \cite{Laird09}.
  In brief, after merging the individual observations into a
single event file, images are constructed in four energy bands
0.5-7.0\,keV (full), 0.5-2.0\,keV (soft), 2.0-7.0\,keV (hard) and
4.0-7.0\,keV (ultra-hard).  The count rates in the above energy
intervals are converted to fluxes in the bands 0.5-10, 0.5-2, 2-10 and
5-10\,keV, respectively.  The limiting flux in each of these bands for
a point source is
estimated to be $3.5 \times 10^{-15}$, $1.1 \times 10^{-16}$, $8.2
\times 10^{-16}$, and $\rm 1.4 \times 10^{-15} \, erg \, s^{-1} \,
cm^{-2}$, respectively. 
Sources are not required to be detected in both the soft and hard bands, 
though most are 
(for the main AGN sample used here, described in Section 3 below, 29\% of
the sources are detected in the soft band only and 9\% are detected 
in the hard band only).
 The X-ray catalog comprises a total of 1325
unique sources over $\rm 0.63\,deg^{2}$ to a Poisson detection
probability threshold of $4 \times 10^{-6}$.  We identify optical
counterparts of the Chandra X-ray sources using the maximum likelihood
(LR) method as described in \cite{Laird09}.  A total of 895
sources have optical counterparts to $R_{\rm AB}=24.1$\,mag ($LR>0.5$).

\subsection{DEEP2 Spectroscopy}

Spectroscopic redshifts were obtained for 463 X-ray sources in AEGIS, 
233 (50\%) of which are from the DEEP2 Galaxy Redshift Survey \citep{Davis00,Davis03} 
and 230 (50\%) of which were obtained with follow-up spectroscopic observations 
at the MMT, as described in \S2.3.  
The DEEP2 survey is a completed project using the
DEIMOS spectrograph \citep{Faber03} on the 10m Keck II telescope to
survey optically-selected galaxies at $z\simeq1$ in a comoving volume of
approximately 5$\times$10$^6$ $h^{-3}$ Mpc$^3$.  Using $\sim1$~hour
exposure times, the survey has measured high-confidence 
redshifts for $>30,000$ galaxies, most in the redshift range $0.7<z<1.5$, 
to a limiting magnitude of $R_{\rm AB}=24.1$.  
%The redshift completeness of 
%the DEEP2 survey is $\sim65$\%, and $\sim65$\% of the galaxies to 
%$R_{\rm AB}=24.1$ were targeted for spectroscopy.
Roughly 65\% of the galaxies to $R_{\rm AB}=24.1$ were targeted for spectroscopy,
and of those targeted, successful redshifts were measured for $\sim$65\%.

One of the DEEP2 fields is the EGS, where the survey has measured
$>11,000$ high-confidence redshifts \citep{Davis07}; unlike other DEEP2 fields,
objects with colors indicating that they have $z<0.7$ are still observed
in the EGS. 
The DEEP2 spectra have high resolution ($R\sim5,000$), and rms redshift
errors (as determined from repeated observations) are $<35$ km
s$^{-1}$.  
After restricting to $0.7<z<1.4$ and the area covered by the AGN samples 
described below, the EGS DEEP2 sample contains a total of 4669 sources.

\subsection{MMT Spectroscopy}

To obtain redshifts for more X-ray AGN than were targeted in the DEEP2
survey (which began before the X-ray observations were taken, so X-ray
sources were not given higher observing priority), we used
the MMT/Hectospec fiber spectrograph for follow-up spectroscopy of
optical counterparts to X-ray
sources.   We observed 5 Hectospec configurations in 
queue mode in May 2007, July 2007, and May 2008. 
Total integration times were 1.5-2 hours per configuration.  The wavelength 
coverage was $\sim4500-9000$ \AA \ at 6 \AA \ resolution. 
Observing conditions were fair, though not ideal;
one night was affected by moonlight, and the seeing was often $>1\arcsec$.
Additional targets were observed as filler objects on configurations taken
by other AEGIS team members (C. Willmer and P. Barmby, priv. comm.) during
the same observing seasons, with similar spectrograph setups.
In total, we targeted optical counterparts for 449 X-ray sources.

The data were reduced using the HSRED IDL reduction 
pipeline\footnote{See http://www.astro.princeton.edu/~rcool/hsred}.
 Redshifts were measured using a $\chi^2$-minimization 
between the observed spectrum and emission-line and absorption-line galaxy 
and AGN templates, and were confirmed by eye. We obtained
high-confidence redshifts for 230 sources, or 51\% of the targeted sample. 

The spatial distribution of the X-ray AGN with redshifts between $0.2<z<1.4$ is 
shown in Figure \ref{skydist} (colored diamonds), 
along with the distribution of DEEP2 galaxies in the same redshift range (black
squares).  As discussed below in Section 3.2, galaxies and AGN in the upper
quadrant are not used for the clustering measurements performed here; the green
squares in Figure \ref{skydist} show the X-ray AGN used to measure the
cross-correlation function.

The $R_{\rm AB}$ magnitudes and $B-R$ colors of the targeted sources and the
subsample with successful redshift measurements are shown in Figure 
\ref{mmt_mags}.  Our success rate declines at $R_{\rm AB}>22.5$, as expected, 
as the signal-to-noise in the 2-hour spectra was not 
high enough to measure a robust redshift.  A K-S test of the $B-R$ color
distribution for the targeted sources and the subsample with successful 
redshifts gives a 93\% probability that they are drawn from the same
parent population.  The success rate is lowest at $B-R\sim1$, but is comparable
for red and blue galaxies.

Figure \ref{mmt_z} shows the redshift distribution 
for all the X-ray sources with spectroscopic redshifts in the EGS (left),
and split according to the instrument used to obtain spectroscopy (right).  
The left panel also shows the redshift distribution of DEEP2 galaxies in the
EGS (dotted line), scaled down by a factor of 35.  The 
redshift distribution for X-ray sources observed with Keck/DEIMOS (right panel)
 shows a peak at 
$z\sim0.7$ and a small tail of objects at $z>1.5$, reflecting both the
underlying redshift distribution of the $R_{\rm AB}<24.1$ DEEP2 sources 
and the high
resolution of the spectra, which results in a narrower spectral range of
 $\sim6500-9000$ \AA, limiting the redshifts where multiple features will be
observed.  The MMT/Hectospec redshift distribution is flatter 
and includes more objects at $z>1.5$, due to the broader wavelength coverage 
of the spectra.  

K-corrections, absolute $M_B$ magnitudes and restframe $(U-B)$ 
colors have been derived as described in \cite{Willmer06}. 
Absolute magnitudes given in this paper are in the AB system 
and are $M_B-5$ ${\rm log}(h)$ with $h=1$, which for the remainder 
of the paper we denote as $M_B$. X-ray luminosities are derived using
$h=0.7$.

%_______________________________________________________________________

\section{AGN Properties and Samples}
\label{sec:agn_samples}

\subsection{Optical Properties of X-ray AGN}

\begin{figure}[t]
\epsscale{1.3}
\plotone{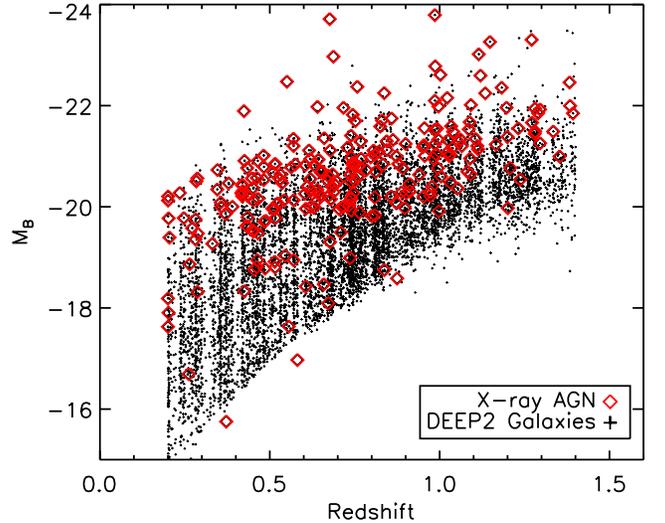}
\caption{Absolute magnitude $M_B$ as a function of redshift for DEEP2 galaxies 
in AEGIS (black crosses) and X-ray AGN (red diamonds).  The AGN typically
reside in optically bright galaxies, though they are occasionally found in 
fainter galaxies.  AGN subsamples as a function of $M_B$ are drawn to test
the luminosity-dependence of AGN clustering.}
\label{mb_z}
\end{figure}

\begin{figure*}[t]
\epsscale{1.1}
\plottwo{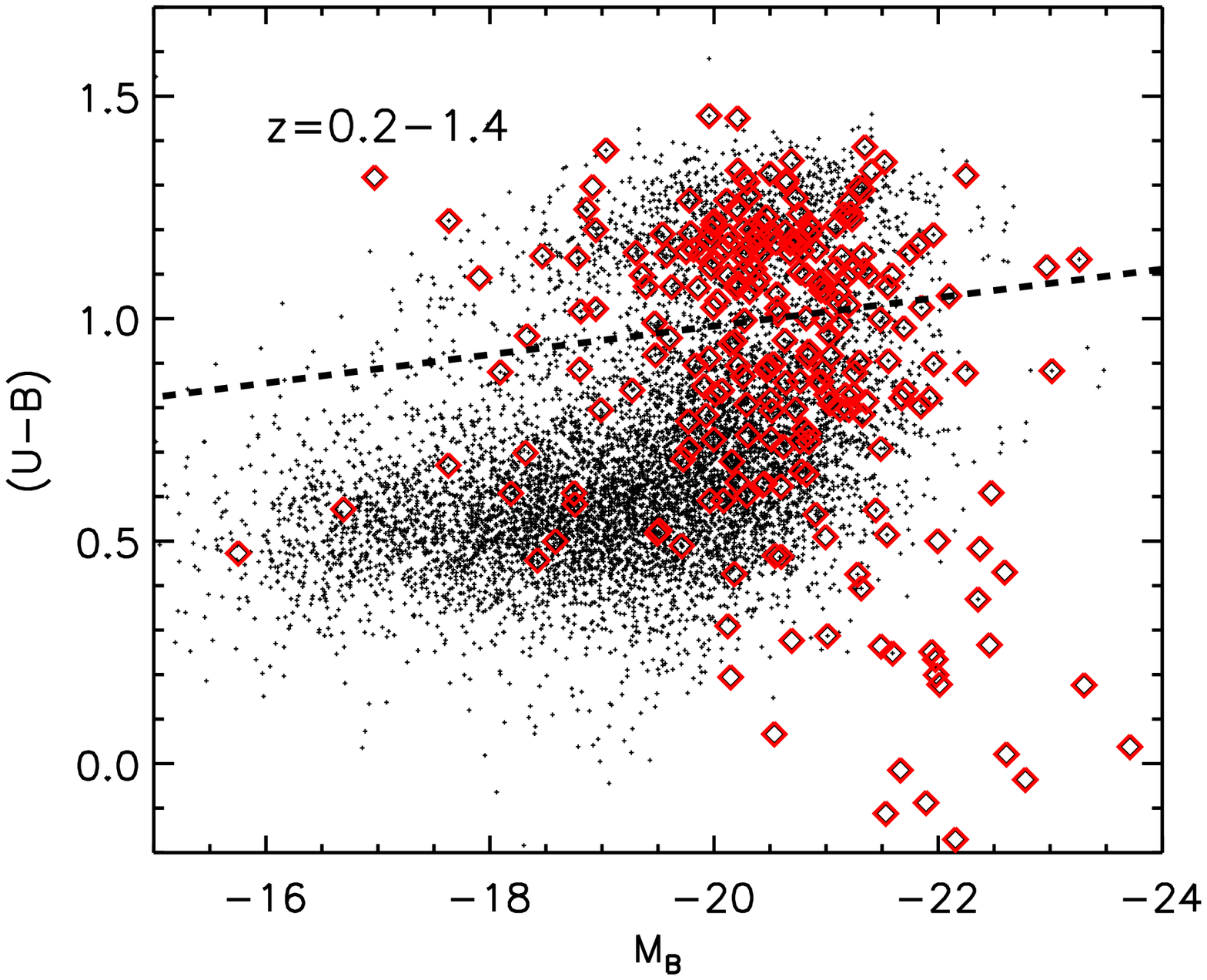}{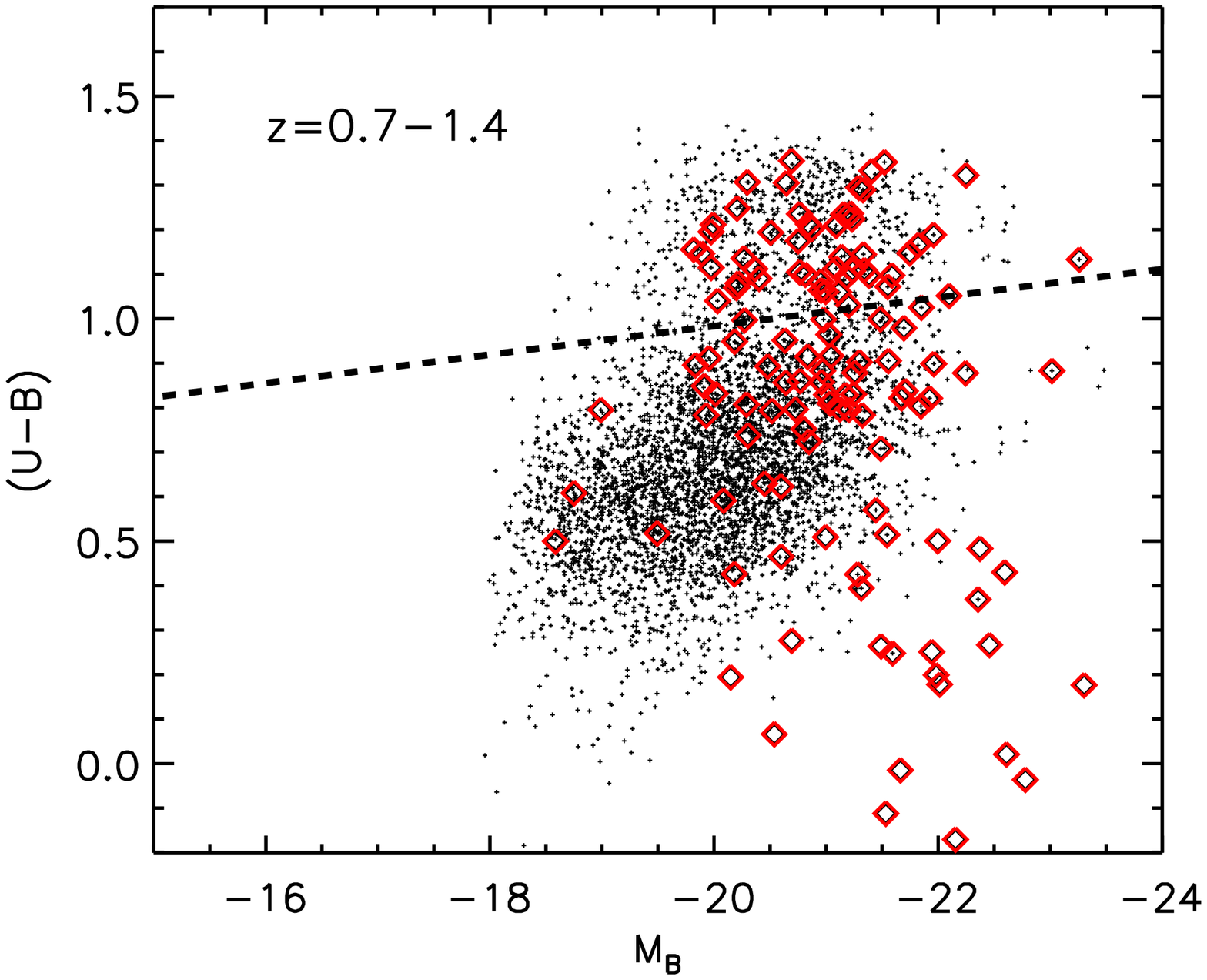}
\caption{Restframe optical color-magnitude ($(U-B)$ versus $M_B$) 
diagram for DEEP2 galaxies (black crosses) and X-ray 
AGN (red diamonds) in AEGIS for the redshift ranges $z=0.2-1.4$ (left) and 
$z=0.7-1.4$ (right). The dashed line shows the definition of the 
minimum in the restframe $(U-B)$ bimodality, used to define red and
blue galaxies. 
 X-ray AGN are predominantly found in red sequence
galaxies, `green' galaxies in the dip of the color bimodality, 
and the brightest/reddest of the blue cloud galaxies.  The
optical light for a small percentage ($\sim14$\%) of the X-ray sources is
dominated by the AGN itself and not the host galaxy; these sources are
seen in the lower right corner and are optically very bright and blue.}
\label{cmd}
\end{figure*}

\begin{figure*}[t]
\epsscale{1.1}
\plottwo{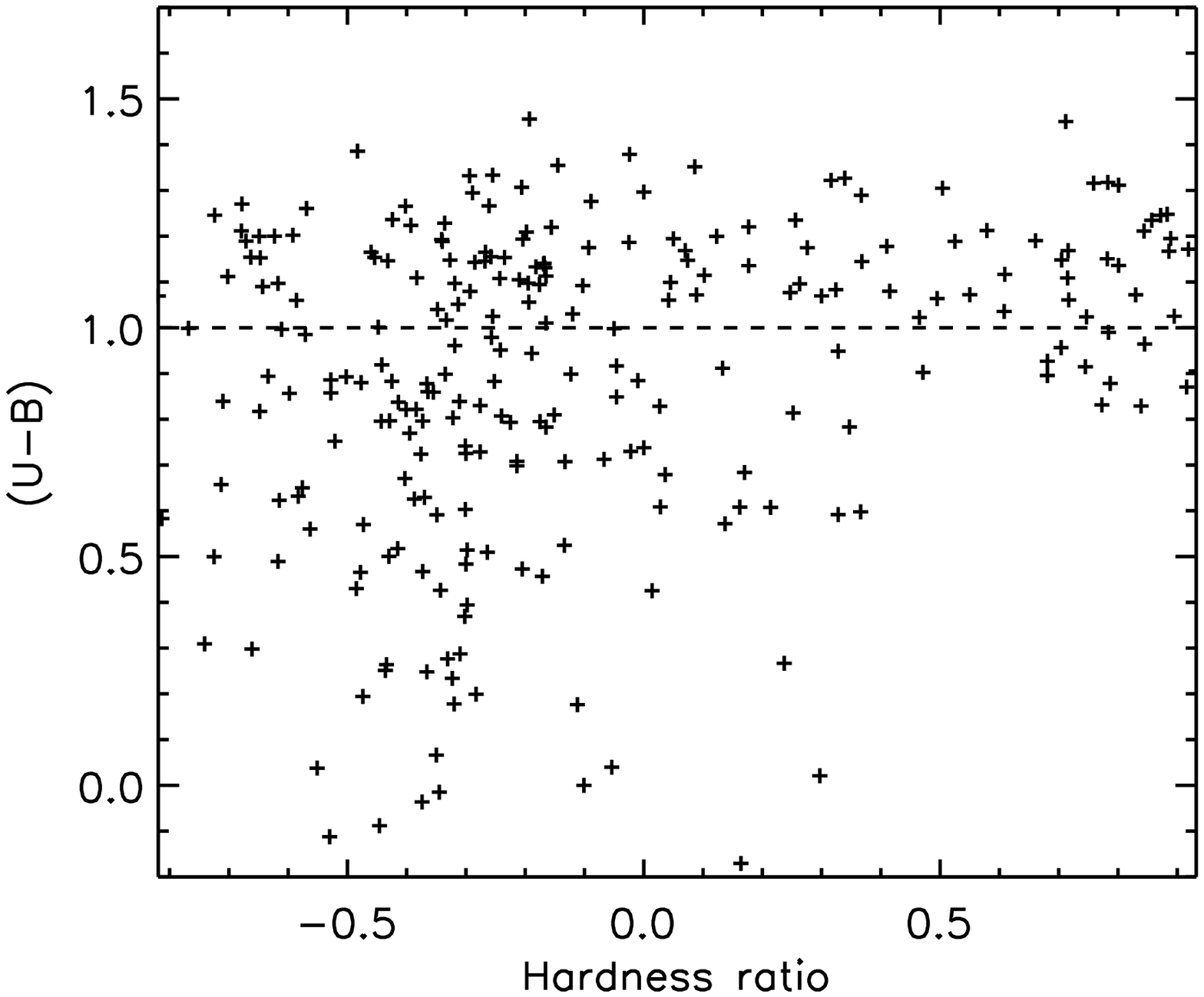}{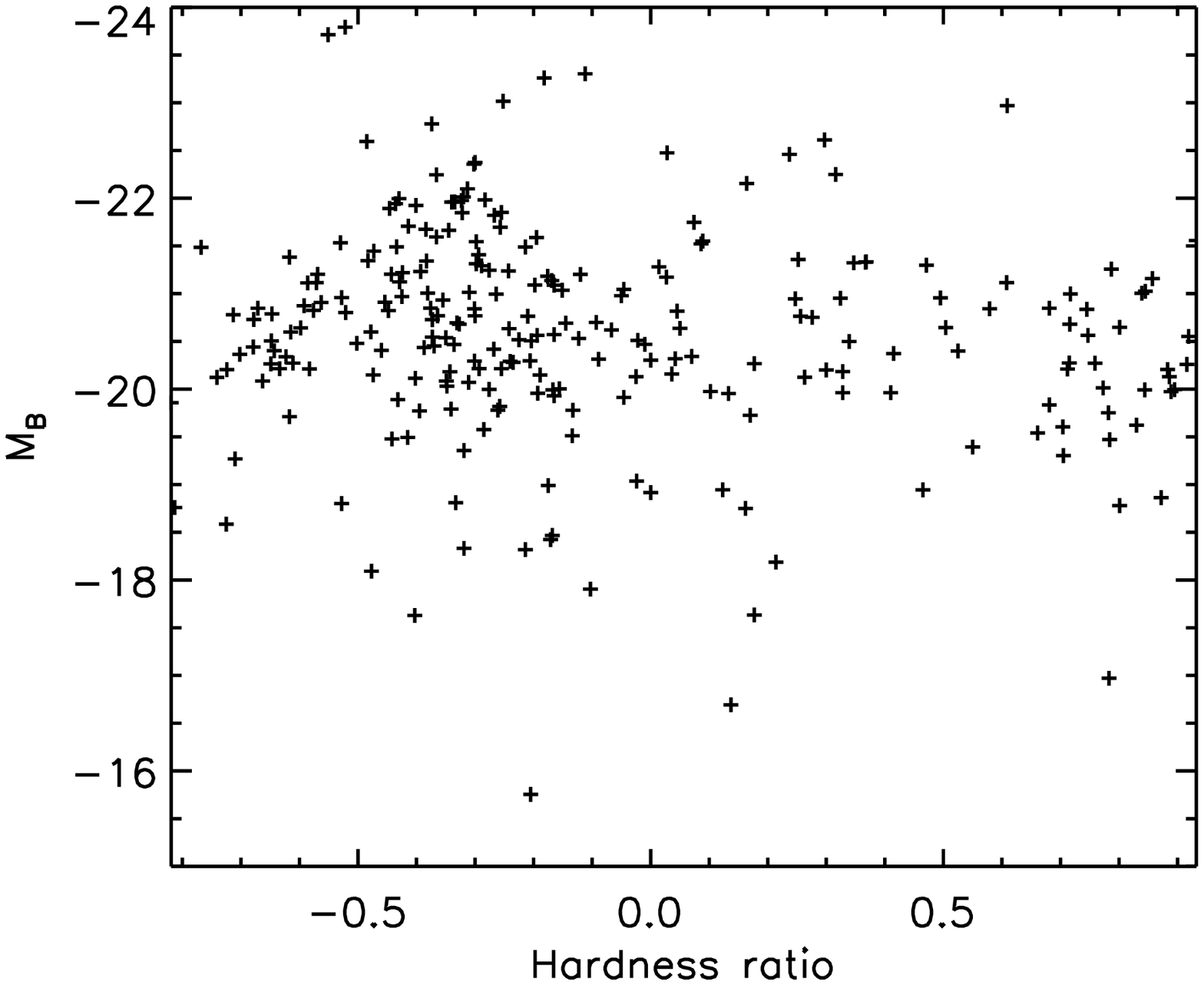}
\caption{Left: Optical $(U-B)$ color as a function of hardness ratio (see 
\S3.1 for details) 
for all X-ray AGN in the redshift range $z=0.2-1.4$. The dashed line roughly
indicates the separation of red and blue galaxies, above and below the line 
respectively.  The optically red and `green' transition galaxies show a 
range of hardness ratios, while the blue galaxies are on average softer 
sources.
Right: Optical absolute magnitude $M_B$ as a function of hardness ratio 
for all X-ray AGN in the redshift range $z=0.2-1.4$. A small but significant
correlation is found between optical magnitude and hardness ratio.}
\label{hardness-optical}
\end{figure*}

\begin{figure*}[t]
\epsscale{1.1}
\plottwo{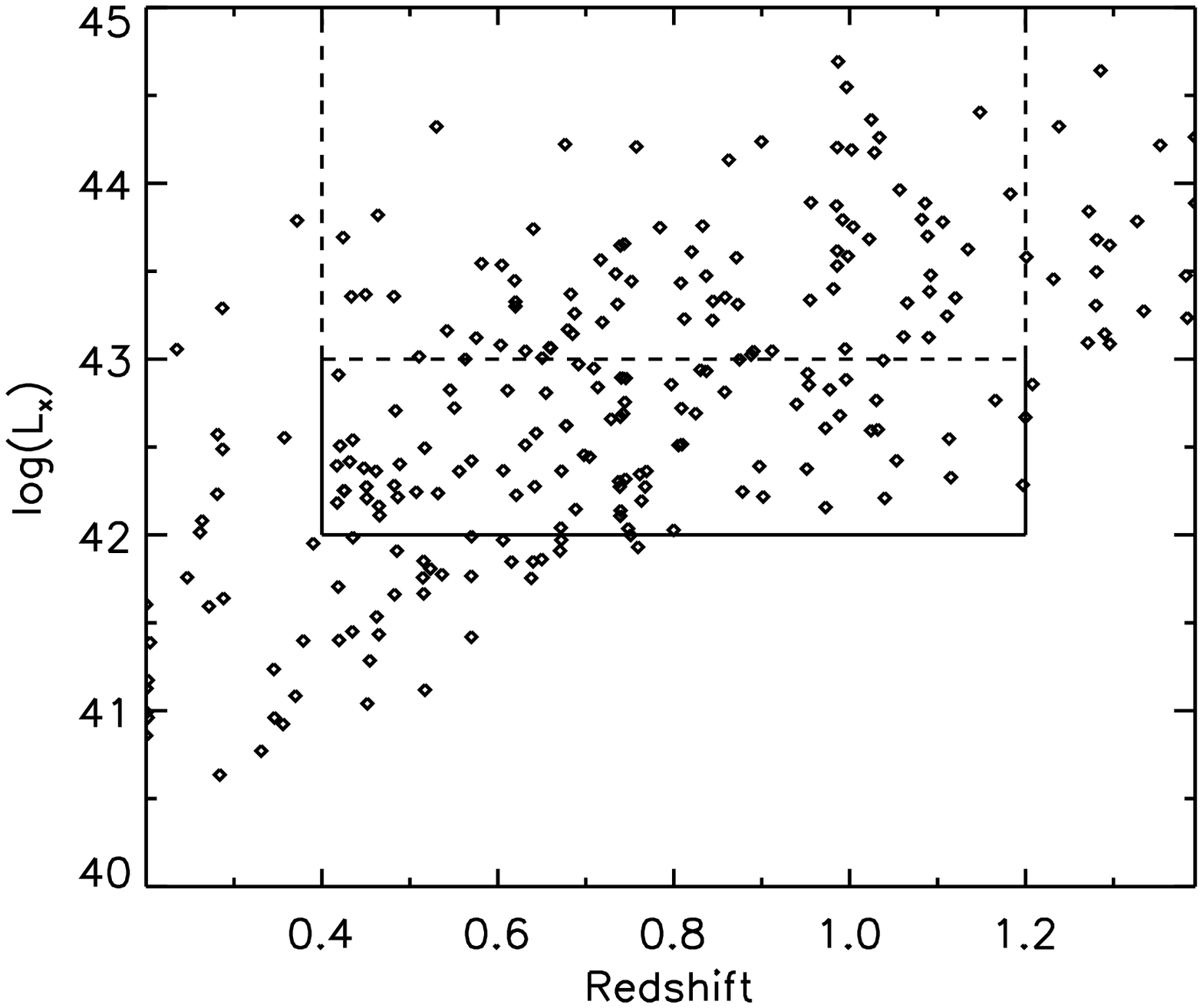}{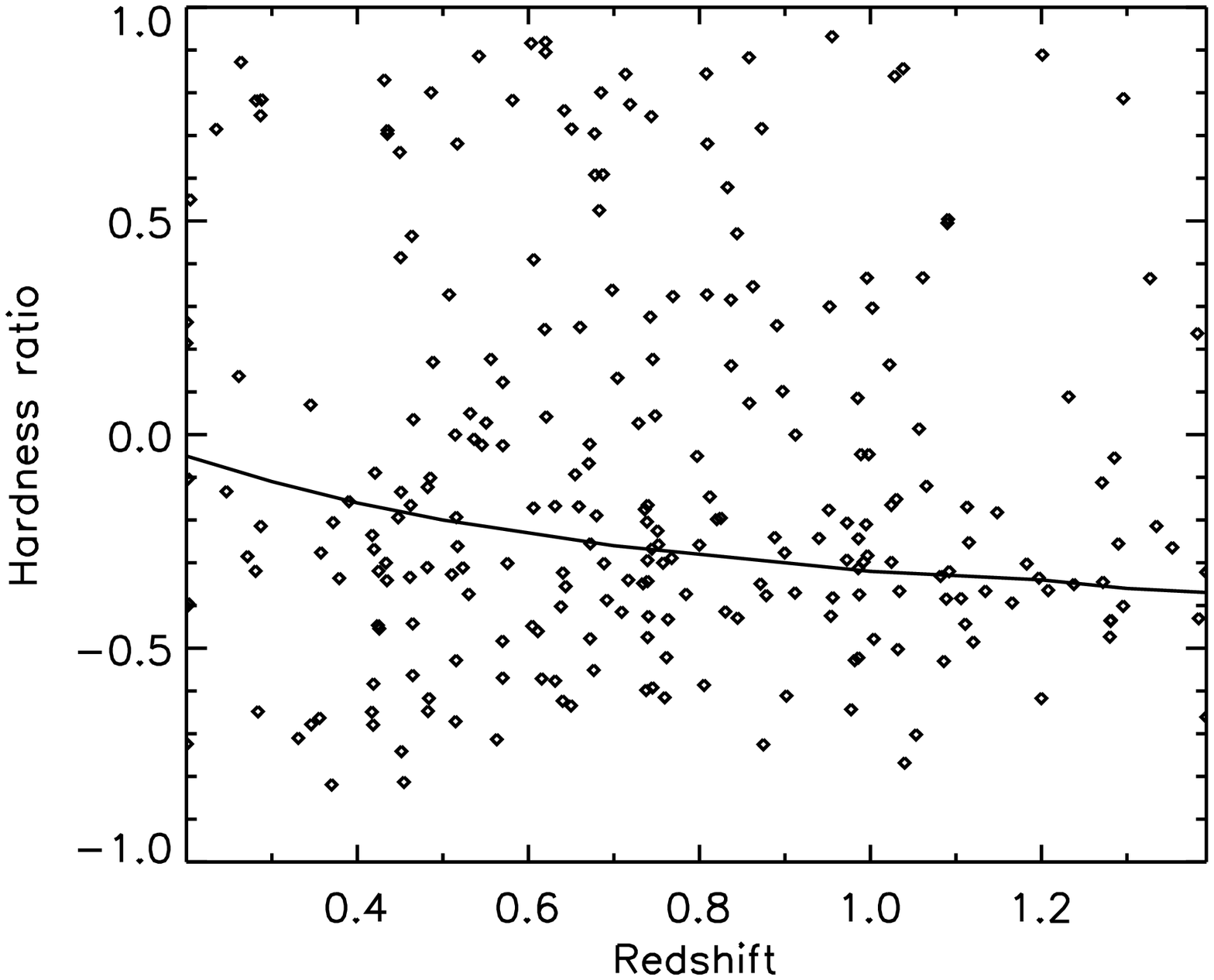}
\caption{Left: X-ray luminosity as a function of redshift for all X-ray AGN in 
the redshift range $z=0.2-1.4$.
The `AGN high $L_{\rm x}$' and `AGN low $L_{\rm x}$' samples are defined by the dashed and 
solid lines, respectively. 
Right: Hardness ratio of the X-ray AGN as a function of redshift.  The `AGN
hard' and `AGN soft' samples are defined as being above or below the solid
line, corresponding to $N_H=10^{22} cm^{-2}$ for $\Gamma=1.9$.}
\label{hardness}
\end{figure*}

Figure \ref{mb_z} shows the absolute magnitude, $M_B$, for the X-ray AGN host
galaxies (red 
diamonds) and all DEEP2 galaxies (black crosses) as a function of redshift.  
The AGN tend to reside in optically bright galaxies, along the upper portion 
of the 
distribution, at all redshifts between $z=0.2-1.4$.  Figure \ref{cmd} shows
the color-magnitude diagram for the X-ray AGN (red diamonds) and DEEP2 
galaxies (black crosses) in the redshift range $z=0.2-1.4$ (left) and 
$z=0.7-1.4$ (right).  As originally shown in \cite{Nandra07} and 
updated here with a larger sample (see also \cite{Barger03}), 
X-ray AGN in the 200-ks depth {\it Chandra} AEGIS data predominantly have host galaxies 
that are either on the red sequence or the bright/red part of the blue
cloud, implying that they reside in massive host galaxies (see also Silverman
et al. 2008)\nocite{Silverman08}.  
X-ray AGN that are very bright and blue,  with $M_B<-21.3$ and $(U-B)<0.3$, 
in the lower right corner of the figure, we define to be quasars, where 
the optical light is dominated by the AGN and not the host galaxy. ACS imaging
of these sources in the EGS shows a bright blue point source with no host
galaxy visible (Pierce et al. 2007, Georgakakis et al. in preparation). 
\nocite{Pierce07}.
14\% of the X-ray AGN in our sample at $0.7<z<1.4$ are quasars, according to this
definition.
We note that contamination of the X-ray AGN sample by normal star-forming
galaxies without AGN in this sample is expected to be less than 
1\% \citep{MonteroDorta08}.

In Figure \ref{hardness-optical} we compare the optical $(U-B)$ color 
with the hardness ratio of the X-ray 
source, measured as $HR\equiv (H-S)/(H+S)$, where $H=2-7$ keV counts
and $S=0.5-2$ keV counts.
The hardness ratio estimates used here are determined through Bayesian 
methods and do not require detection in both the hard and soft bands; the 
actual source counts and background counts in both bands are used even if 
the object is not significantly detected.  Therefore none of the X-ray AGN 
have $HR=-1$ or $HR=+1$.
The dashed line roughly marks the minimum of the color bimodality, with
blue galaxies at $(U-B)<1.0$ and red galaxies at $(U-B)>1.0$. 
We find a correlation between hardness ratio and optical
obscuration in that the hardest X-ray sources are all 
optically red  \citep[e.g.,][]{Mainieri02}.  This may either be the 
result of obscuration in the host galaxy itself, which reddens the galaxy 
as well as obscures the AGN, or it may be 
that obscured AGN only reside in galaxies with older stellar populations
\citep{Nandra07,Rovilos07}.
We find that the red and `green' galaxies have a wide range of hardness ratios, 
while the blue galaxies are softer sources.  A small but significant 
correlation 
is found between absolute magnitude $M_B$ and hardness ratio, which have
a Spearman's rank correlation coefficient of $\rho=0.14$ at 97\% significance.
This correlation persists after excluding the brightest sources with $M_B<-22$,
indicating that it is not likely to be dominated by an increased contribution
from unobscured AGN.

\begin{deluxetable*}{lrcccc}[t]
\tablewidth{0pt}
\tablecaption{AGN and Galaxy Samples}
\tablehead{
\colhead{Sample}&\colhead{No. of} &\colhead{$z$}  &\colhead{median}&\colhead{median}&\colhead{selection\tablenotemark{a}}\\
\colhead{}      &\colhead{objects}&\colhead{range}&\colhead{$z$} &\colhead{$M_B$} &\colhead{}
}
\startdata
AGN main & 113 & 0.7-1.4 & 0.90 & -20.98 &  does not include quasars \\ 
AGN host bright & 54 & 0.7-1.4 & 1.00 & -21.33 & $M_B<-21$ \\ % roughly volume-limited
AGN host faint  & 59 & 0.7-1.4 & 0.84 & -20.45 & $M_B>-21$ \\ % roughly volume-limited
AGN host red  & 50 & 0.7-1.4 & 0.90 & -20.96 & red $(U-B)$ host galaxy \\ % $(U-B)>-0.032 (M_B+21.62)+1.035$ \\
AGN host blue & 63 & 0.7-1.4 & 0.91 & -20.98 & blue $(U-B)$ host galaxy \\ %$(U-B)<-0.032 (M_B+21.62)+1.035$ \\
AGN low $L_{\rm x}$  & 91 & 0.4-1.2 & 0.74 & -20.40 & $42<{\rm log} L_{\rm X}<43$ erg $s^{-1}$ \\
AGN high $L_{\rm x}$ & 83 & 0.4-1.2 & 0.84 & -21.01 & log $L_{\rm X}>43$ erg $s^{-1}$ \\
AGN hard & 140 & 0.2-1.4 & 0.75 & -20.62 & $N_H>10^{22} cm^{-2}$ \\
AGN soft & 111 & 0.2-1.4 & 0.67 & -20.60 & $N_H<10^{22} cm^{-2}$ \\

All galaxies  & 4669 & 0.7-1.4 & 0.92 & -20.16 & \\
Red galaxies  &  757 & 0.7-1.4 & 0.85 & -20.66 & red $(U-B)$ \\
Blue galaxies & 3913 & 0.7-1.4 & 0.94 & -20.06 & blue $(U-B)$ \\
Matched galaxies      & 504 & 0.7-1.4 & 0.98 & -20.97 &  \\
Red matched galaxies  & 226 & 0.7-1.4 & 0.92 & -20.92 & \\
Blue matched galaxies & 278 & 0.7-1.4 & 1.03 & -20.99 & \\

\enddata
\tablenotetext{a}{See Section 3 for details on each AGN sample and Section 5 for details on each galaxy sample.}
\label{samples}
\end{deluxetable*}

\subsection{AGN samples}

From the full AGN and galaxy datasets we define several samples used to
measure the AGN-galaxy cross-correlation function.
We limit most samples to the redshift range
$0.7<z<1.4$ so as to match the redshift range of the other DEEP2 fields where
we have performed extensive clustering analyses of galaxies as a function
of color and luminosity \citep{Coil06lum,Coil08}, as well as the
measurement of the quasar-galaxy
cross-correlation function \citep{Coil07}, with which we compare our results
here.  We do not include data 
from the upper quadrant of the EGS field (pointing 14 in the DEEP2 
naming convention), where the DEEP2 target selection function is slightly
different from the rest of the survey, due to shallower $BRI$ photometric 
imaging \citep{Coil04} used in the target selection. The total area covered
by the data used here is $\sim0.4$ deg$^2$. 

Details of each AGN sample are given in Table \ref{samples}.
The `AGN main' sample includes all non-quasar X-ray AGN over the redshift range
$z=0.7-1.4$, a total of 113 sources; as seen in the left panel of 
Figure \ref{hardness} these AGN have log $L_{\rm X}>42$ erg $s^{-1}$, where 
$L_{\rm X}$ is the restframe 2-10 keV luminosity calculated from 
the observed 2-10 keV flux, assuming a photon index $\Gamma=1.9$.

The main AGN sample has a 
mean log $L_{\rm X}=42.4$ erg $s^{-1}$ and a median log $L_{\rm X}=42.8$ erg $s^{-1}$.
The `AGN host bright' and `AGN host faint'
samples are defined by their optical $M_B$ magnitude; the samples are
split at the median $M_B$ magnitude of the AGN main sample. The `AGN host red' 
and `AGN host blue' samples are defined by the optical color of the host 
galaxy.  Following \cite{Willmer06}, we define red and blue
galaxies using the observed color bimodality in 
DEEP2 with the following tilted cut in color-magnitude space 
(in AB magnitudes):
\begin{equation}
(U-B)=-0.032 (M_B+21.62)+1.035,
\end{equation}
as shown in Figure \ref{cmd}.
We do not allow this color cut to evolve with redshift within the 
redshift range used here.  

For the AGN samples defined by optical color or
 magnitude we exclude the quasars 
in the lower right portion of Figure \ref{cmd}, as the optical color
and magnitude of these objects is not reflective of the host galaxy but is
from the AGN itself.  We define quasars here as sources with 
$M_B<-21.3$ and $(U-B)<0.3$; there are 19 quasars in our sample with 
$0.7<z<1.4$ (12 of which are shown in the right panel of Figure \ref{cmd}, 
the rest are outside the plot range), out of 132 X-ray sources. None of our results change if 
these sources are included. 
For the `AGN host bright/faint' and `AGN host red/blue' samples, 
the optical magnitude and color are properties of the host galaxy, 
not the AGN itself. The sample of spectroscopically-identified broad-line AGN
is too small to use for robustly measure clustering properties; a larger sample
is needed.

We also divide the full AGN sample by X-ray luminosity, using the median 
value of log $L_{\rm x}=43$ erg $s^{-1}$, such that the 
`AGN low $L_{\rm x}$' sample is defined as having $42<$log $L_{\rm X}<43$ erg $s^{-1}$ 
(which results in a roughly volume-limited sample),
while the `AGN high $L_{\rm X}$' sample has log $L_{\rm X}>43$ erg $s^{-1}$, as 
shown in the left panel of Figure \ref{hardness}.  
Since here we are concerned
with comparing the $L_{\rm X}$ samples to each other and not to galaxies in
the same redshift range,  we are free to use a different redshift range
from that used in the DEEP2 galaxy clustering studies. 
For the $L_{\rm X}$-selected samples we use a lower redshift limit of $z=0.4$,
 which increases the sample size but is not so low as to cause the two 
samples to have significantly different mean redshifts (see Table 1). 
To keep the `AGN low $L_{\rm X}$' sample roughly 
volume-limited we use an upper redshift limit of $z=1.2$ 
(see Figure \ref{hardness}), which ensures that any difference seen
in their clustering properties should be dominated by differences in 
luminosity and not redshift.  
The `AGN hard' and `AGN soft' samples are defined
by the X-ray color, or hardness ratio, as shown in the right panel of Figure
\ref{hardness}.  The solid line corresponds to $N_H=10^{22} cm^{-2}$ for 
$\Gamma=1.9$ and defines the `AGN hard' and `AGN soft' samples.  For these 
samples we use the full redshift range of $z=0.2-1.4$.

Several galaxy samples are used in this paper; details for each are given
in Table 1.  All cross-correlation functions
presented here use the full DEEP2 galaxy sample with $0.7<z<1.4$ (`All galaxies'
in Table 1) to cross-correlate with either an AGN or a galaxy sample.  This
full galaxy sample is separated into red and blue galaxy samples based on
restframe $(U-B)$ color, with the color cut given above. Matched galaxy samples
are discussed in \S5.3; these samples have the same $M_B$ and $(U-B)$ distributions
as various comparison AGN samples.

%_______________________________________________________________________

\section{Measuring the Cross-Correlation Function}
\label{sec:cross}

To measure the clustering of the X-ray AGN sample, we use the cross-correlation
of AGN with galaxies in the same volume, and the auto-correlation function
of the galaxies themselves, to infer the auto-correlation function of
the AGN sample.

The two-point auto-correlation function \xir \ is defined as a measure
of the excess probability above that for an unclustered distribution 
of finding an object in a
volume element $dV$ at a separation $r$ from another randomly chosen
object,
\begin{equation}
dP = n [1+\xi(r)] dV,
\end{equation}
where $n$ is the mean number density of the object in question
\citep{Peebles80}.  The cross-correlation function is the excess
probability above Poisson of finding an object from a given sample at
a separation $r$ from a random object in another sample.  Here we
measure the cross-correlation between AGN and galaxies:
\begin{equation}
dP(G|A) = n_G [1+\xi_{AG}(r)] dV,
\end{equation}
which is the probability of finding a galaxy ($G$) in a volume element
$dV$ at a separation $r$ from an AGN ($A$), where $n_G$ is the
number density of galaxies.

To estimate the cross-correlation function between our AGN and
galaxy samples, we measure the observed number of galaxies around each
AGN as a function of distance, divided by the expected number of
galaxies for a random distribution.  We use the estimator 
\begin{equation}
\xi=\frac{AG}{AR}-1,
\end{equation}
where $AG$ are AGN-galaxy pairs and $AR$ are AGN-random pairs at
a given separation, where the pair counts have been normalized by
$n_G$ and $n_R$, respectively, the mean number densities in the full galaxy
and random catalogs.  This estimator is preferred here as it does not
require knowledge of the AGN selection function, only the galaxy
selection function, which is well-quantified.  To calculate \xir \ 
 we create a random catalog with
the same redshift distribution as the DEEP2 galaxies used for the
cross-correlation measurement and the same 
sky coverage as the DEEP2 galaxies in the EGS, applying
the two-dimensional variation in the DEEP2 target selection and
redshift success rate over the plane of the sky.
We also mask the regions of the random
catalog where the photometric data are affected by saturated stars or CCD
defects.  Details on the random catalog construction are given in \S3.1 of 
\cite{Coil04xisp}. 

We wish to measure the real-space correlation  
function, \xir.  However, peculiar velocities distort the positions  
of objects in redshift space along the line of sight.  To capture  
this effect, we measure $\xi$ as a function of two coordinates,
perpendicular to ($r_p$) and along ($\pi$) the line of sight. 
In applying the above estimator to galaxies, pair counts are
computed over a two-dimensional grid of separations to estimate \xisp.
To recover \xir, \xisp \ is integrated along the $\pi$ direction and 
projected onto the $r_p$ axis.  As
redshift-space distortions affect only the line-of-sight component of
\xisp, integrating over the $\pi$ direction leads to a statistic
\wprp, which is independent of redshift-space distortions.  Following
\cite{Davis83},
\begin{equation}
w_p(r_p)=2 \int_{0}^{\infty} d\pi \ \xi(r_p,\pi)=2 \int_{0}^{\infty}
dy \ \xi[(r_p^2+y^2)^{1/2}],
\label{eqn}
\end{equation}
where $y$ is the real-space separation along the line of sight. 
Here, as in other DEEP2 clustering studies, 
we integrate to a maximum separation in the $\pi$ 
direction of 20 \mpch, as the signal to noise degrades quickly for larger 
separations where $\xi$ becomes small.  
Systematic effects due to the use of slitmasks are small (see \S3.4 of 
\cite{Coil08} for details).  We do not correct for these effects here, as we
are primarily concerned with the {\it relative} clustering of AGN samples 
compared to each other and to the clustering of galaxy samples, and to 
first order these effects will cancel.

\begin{deluxetable*}{lcc}[t]
%\tabletypesize{\scriptsize}
\tablewidth{0pt}
\tablecaption{Relative Bias Results}
\tablehead{
\colhead{Samples}&\colhead{Relative bias}&\colhead{Relative bias}\\
\colhead{}&\multicolumn{1}{c}{$0.1<r_p<8 \ h^{-1}$ Mpc}&\multicolumn{1}{c}{$1<r_p<8 \ h^{-1}$ Mpc} \\
}
\startdata
AGN host bright/faint & $1.52 \pm0.32$ & $1.20 \pm0.31$ \\
AGN host red/blue & $1.91 \pm0.38$ & $1.79 \pm0.27$ \\
AGN high $L_{\rm x}$/low $L_{\rm x}$ & $1.05 \pm0.18$ & $0.95 \pm0.20$ \\
AGN hard/soft & $1.22 \pm0.12$ & $1.27 \pm0.17$ \\
AGN/Red galaxies & $0.97 \pm0.07$ & $0.94 \pm0.08$ \\
AGN/Blue galaxies & $1.61 \pm0.11$ & $1.48 \pm0.12$ \\
AGN/Matched galaxies & $1.28 \pm0.10$ & $1.15 \pm0.09$ \\
AGN host red/Red galaxies & $1.20 \pm0.09$ & $1.13 \pm0.09$ \\
AGN host red/Red matched galaxies & $1.33 \pm0.15$ & $1.19 \pm0.12$ \\
AGN host blue/Blue galaxies & $1.27 \pm0.15$ & $1.22 \pm0.13$ \\
AGN host blue/Blue matched galaxies & $1.24 \pm0.16$ & $1.09 \pm0.16$ \\

\enddata
%\tablenotetext{a}{}
\label{relbias}
\end{deluxetable*}

Error bars on \wprp \ are estimated from jackknife sampling of the data, 
where the EGS is divided into eight regions of equal area and \wprp \ is 
calculated for eight jackknife samples, removing one area at a time.  To check
the validity of these jackknife error bars, we compare the \wprp \ 
errors for the `main AGN sample' with errors that include both 
Poisson and cosmic variance errors estimated using 
 DEEP2 mock catalogs derived 
from the Millenium Run simulation \citep[see][for further details]{Manfred06}. 
Using a total of 24 independent mock catalogs with the same geometry and
selection function as the DEEP2 data in the EGS, we select 113 random 
red galaxies with $0.7<z<1.4$ to be X-ray AGN proxies, since red DEEP2 
galaxies have a similar clustering amplitude as the X-ray AGN, as shown below 
in Section 5.3.
We calculate the cross-correlation of these AGN proxy galaxies
with all mock DEEP2-like galaxies in each mock catalog, and then
calculate the variance across the 24 catalogs.  The error on \wprp \ 
in the mock catalogs is less than the error estimated from jackknife sampling 
of the data itself: on scales $r_p$=1-8 \mpch \ in the mock catalogs the
fractional error on \wprp \ is 8\%, while in the data it is 10\%.  To be
conservative we therefore use
the error estimated from the jackknife samples of the data. As previously noted 
\citep{Coil07}, the Millenium Run mock catalogs appear to exhibit smaller 
variance than the real Universe.  

If \xir \ is modeled as a power law, $\xi(r)=(r/r_0)^{-\gamma}$, then \rr \ 
and $\gamma$ can be readily extracted from the projected correlation
function, \wprp, using an analytic solution to Equation \ref{eqn}:
\begin{equation}
w_p(r_p)=r_p \left(\frac{r_0}{r_p}\right)^\gamma
\frac{\Gamma(\frac{1}{2})\Gamma(\frac{\gamma-1}{2})}{\Gamma(\frac{\gamma}{2})},
\label{powerlawwprp}
\end{equation}
where $\Gamma$ is the usual gamma function.  A power-law fit to \wprp \
will then recover \rr \ and $\gamma$ for the real-space correlation
function, \xir.  
In practice, however, we can not measure \xisp \ accurately to 
infinite separations as assumed in Equation \ref{eqn}, and here we 
integrate \wprp \ to $\pi_{\rm max}=20$ \mpch. To recover the
correlation length and slope we then estimate \wprp \ integrated to 
$\pi_{max}=20$ \mpch \ for a grid of \rr \ and $\gamma$ values and minimize 
the difference between the data and model.  Full details are given in \S4.2 
of \cite{Coil08}.  We do not use a full covariance matrix in the fits.  
Errors on \rr \ and $\gamma$ are propagated from the jackknife errors on \wprp; 
if instead we measure \rr \ and $\gamma$ in the separate jackknife samples and 
use the variance among the various fits, which does take into account the 
covariance between the $r_p$ bins, the resulting errors are somewhat smaller.  
To be conservative we use the larger errors derived from the \wprp \ errors.

%_______________________________________________________________________

\section{Results}

\begin{figure}[t]
\epsscale{1.1}
\plotone{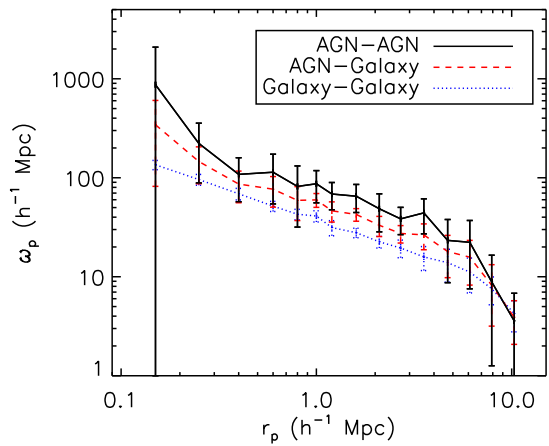}
\caption{
The projected AGN-galaxy cross-correlation function \wprp \ 
as a function of scale for the non-quasar X-ray AGN main 
sample (red dashed line) and the full DEEP2 galaxy population in the EGS 
with $0.7<z<1.4$.  The galaxy auto-correlation function in the same
volume is shown (blue dotted line), as well as the inferred X-ray AGN
auto-correlation function (black solid line). 
Errors are computed from jackknife samples of the data and 
are comparable to  errors found from mock catalogs
(see text for details). 
 }
\label{main}
\end{figure}

\begin{figure*}[t]
\epsscale{1.1}
\plotone{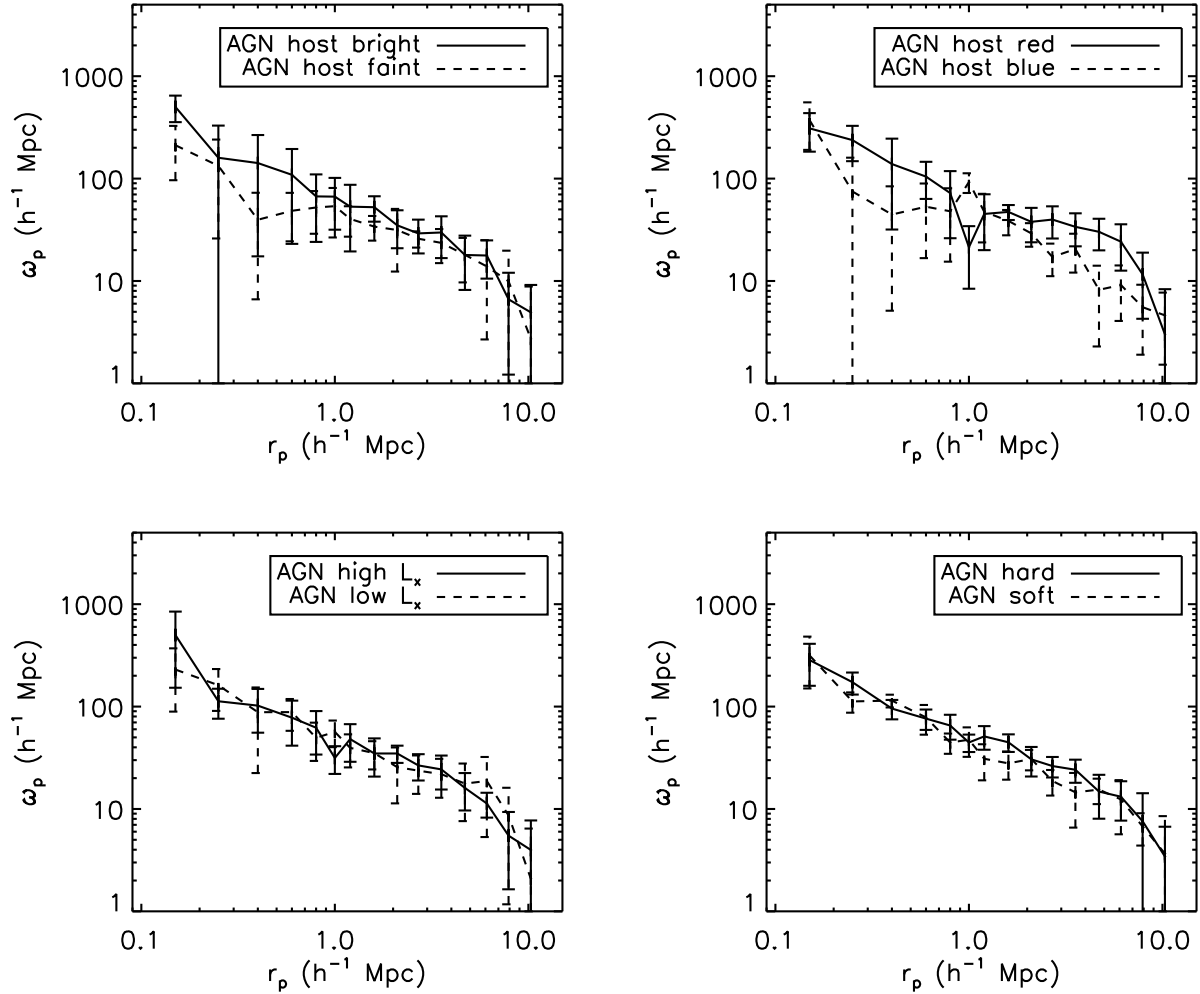}
\caption{The projected AGN-galaxy cross-correlation functions for 
AGN samples split according to optical $M_B$ magnitude (upper left), 
optical $(U-B)$ color (upper right), X-ray luminosity (lower left), and
X-ray color or hardness ratio (lower right).  Details of each sample
are given in Table 1.  No significant difference in
clustering is detected for the samples selected on optical luminosity, X-ray 
luminosity or X-ray color.  There is a correlation with optical color: the 
X-ray AGN with red host galaxies are significantly more clustered than those 
in blue host galaxies. }
\label{AGNsamples}
\end{figure*}

\subsection{Clustering of Main AGN Sample}

The projected cross-correlation of the `AGN main' sample with all DEEP2 galaxies 
in the same redshift range, \wprp, is shown with a dashed red line in Figure \ref{main}.
As discussed above, the \wprp \ error bars are derived from jackknife samples
of the data and are slightly larger than errors inferred from cosmological 
simulations.  The projected galaxy auto-correlation function is also shown (blue
dashed line) for the full galaxy sample with $0.7<z<1.4$.  
Assuming a linear bias, 
\begin{equation}
w_p(AG) = [w_p(AA) \times w_p(GG)]^{1/2},
\end{equation}
where $w_p(AG)$ is the AGN-galaxy cross-correlation function, and $w_p(AA)$ 
and $w_P(GG)$ are the AGN and galaxy auto-correlation functions.  From 
measurements of $w_P(AG)$ and $w_P(GG)$, as shown in Figure \ref{main}, 
the X-ray AGN auto-correlation function $w_P(AA)$ can then be inferred; 
this is shown in Figure \ref{main} as a solid black line,
where the errors are derived from jackknife resampling of $w_P(AG)^2/w_P(GG)$.

Fitting $w_p(AA)$ as a power law, we find $r_0=5.95 \pm0.90$ \mpch \ 
and $\gamma=1.66 \pm0.22$.
To compute the bias of the X-ray AGN auto-correlation 
function compared to that expected for dark matter, we estimate \wprp \ for dark 
matter particles at the mean redshift 
of the main AGN sample, using the publicly-available code of \cite{Smith03}, 
where we integrate the dark matter \xisp \ to $\pi_{\rm max}=20$ \mpch.  
We  estimate the linear galaxy bias from the ratio of the quantities 
$b^2=[w_p]_{\rm gal}/[w_p]_{\rm dark matter}$, and average over scales
$1<r_p<8$ \mpch,  to find a mean bias of $b=1.85 \pm0.28$.

From the bias of the main AGN sample we can
infer the minimum dark matter mass for halos that host these AGN.
As shown in the Appendix of \cite{Zheng07}, the dark matter mass inferred
assuming that all objects, whether galaxies or AGN, are `central' objects
in their dark matter halos (ie., assuming one galaxy or AGN per halo) can 
lead to an overestimate of the minimum halo mass.  Here we
assume that the AGN population has the same satellite fraction, or number
of additional galaxies per halo, as found in the DEEP2 galaxy population using
HOD modeling of the correlation function, where the satellite fraction is 
$\sim$15\% \citep{Zheng07}.  Using
this correction for the satellite fraction changes the relative weights given
to halos of a different mass when computing the large-scale bias for different
halo masses, and the minimum mass inferred is slightly lower
than assuming one AGN per halo.  Here 
we find that at $z=0.94$ the minimum dark matter halo mass of the non-quasar
X-ray AGN is $M_{min}=5 \ (+5/-3) \times 10^{12} h^{-1} M_{\sun}$, similar to 
red galaxies at the same redshift \citep{Coil08}.

\begin{figure}[t]
\epsscale{1.2}
\plotone{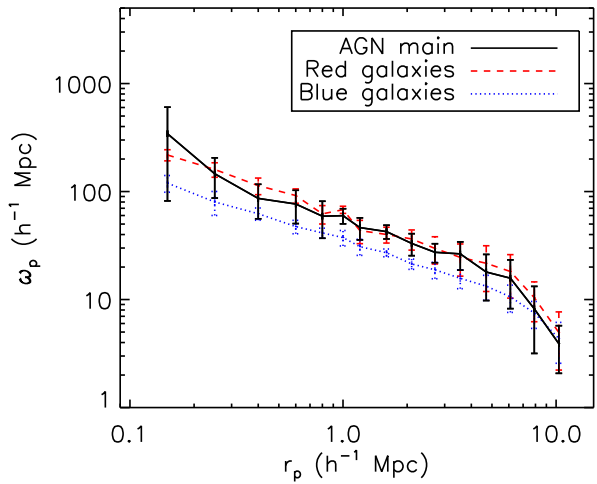}
\caption{The projected AGN-galaxy cross-correlation function \wprp \ 
as a function of scale for the AGN main 
sample (black solid line) and the full DEEP2 galaxy sample
with $0.7<z<1.4$.  The cross-correlations of red and blue DEEP2 
galaxies with the full galaxy sample 
in the same volume are also shown with dashed red and dotted 
blue lines, respectively.  The X-ray AGN are seen to have very similar 
clustering properties to red galaxies and are more clustered than blue galaxies 
at $z\sim1$.
}
\label{gals}
\end{figure}

\subsection{Clustering of Other AGN Samples}

To investigate the dependencies of AGN clustering properties, we divide the 
main AGN sample into samples based on optical luminosity,
optical color, X-ray luminosity and hardness ratio.  Details of the 
selection for each sample are given in Section \ref{sec:agn_samples} 
and in Table \ref{samples}.
Figure \ref{AGNsamples} shows the projected cross-correlation with 
the DEEP2 galaxies, \wprp, for each of these samples.  
To quantify differences in clustering amplitude, we calculate the 
relative bias between these samples.  The relative bias is defined
here as the ratio of \wprp \ for two samples.  Note that this is 
the usual meaning of relative bias ($b_{rel}^2=\xi_1/\xi_2$), 
as here we are using cross-correlation functions with the identical galaxy tracer
samples and not auto-correlation
functions.  The error bars on the relative bias measurement are not 
estimated from the errors on \wprp \ directly, as that would not include 
covariance between the bins and would include cosmic variance errors, 
which cancel to first order when comparing the clustering of two populations
in the same volume.  The relative bias errors are therefore estimated from 
the variance of the relative bias measured across the jackknife samples.

As the relative bias can be scale-dependent,
we calculate the unweighted mean relative bias over two scales, $r_p=0.1-8$ \mpch
\ (`all scales') and $r_p=1-8$ \mpch \ (`large scales'), and also measure
the variance of the relative bias on the same scales.
Relative bias results for various samples are given in Table \ref{relbias}.
We find no significant difference in the clustering of AGN when selected
by optical luminosity, X-ray luminosity, or hardness ratio, within the ranges
sampled here.  We find that the AGN in red host galaxies 
are significantly more clustered than those in blue host 
galaxies, at the 2.9$\sigma$ level 
on large scales, with a relative bias of $b_{rel}=1.79 \pm0.27$, and at the
2.4$\sigma$ level on all scales, with a relative bias of 
$b_{rel}=1.91 \pm0.38$.  This relative
bias is consistent with the relative bias found between red and blue DEEP2
galaxies of $b_{rel}=1.44 \pm0.07$ measured in \cite{Coil08}.

We do not find a significant difference in the clustering of hard X-ray
sources compared to soft X-ray sources.  This lack of a detectable correlation
between hardness ratio and clustering amplitude may be surprising, given that 
we observe correlations both between host $(U-B)$ color and hardness ratio 
(in that the hardest sources are in optically red host galaxies, see Figure 
\ref{hardness-optical}) and between host $(U-B)$ color and clustering amplitude,
both using smaller samples than for the hardness ratio-clustering comparison.
One reason for the lack of a significant 
correlation between hardness ratio and clustering
amplitude may be the large range of hardness ratios among the red AGN
host galaxies
(see Figure \ref{hardness-optical}), so that while the 
hard sample is dominated by red galaxies, the soft sample has both red 
and blue galaxies, mitigating the effect of the hardness-color and color- 
clustering correlations.
Our results clearly show that AGN host color is 
more correlated with dark matter halo mass than hardness ratio is: 
i.e., star formation histories are more closely  
related to local environment than AGN X-ray hardness is.

\begin{figure}[t]
\epsscale{1.2}
\plotone{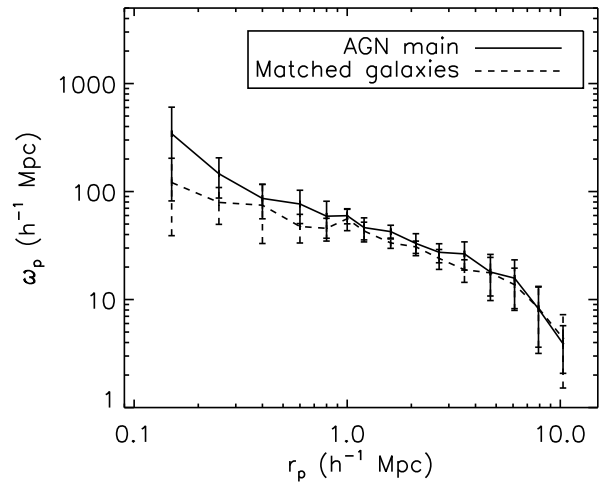}
\caption{The projected AGN-galaxy cross-correlation function \wprp \ for 
the AGN main sample compared to the cross-correlation function of a sample of galaxies 
with the same optical color and magnitude distribution, regardless of whether
they host AGN.  Those galaxies with X-ray AGN are found to cluster more than the full galaxy population with the same color and luminosity distribution, which 
implies that they reside in more massive dark matter halos.
}
\label{all_gals}
\end{figure}

\begin{figure*}[t]
\epsscale{1.1}
\plotone{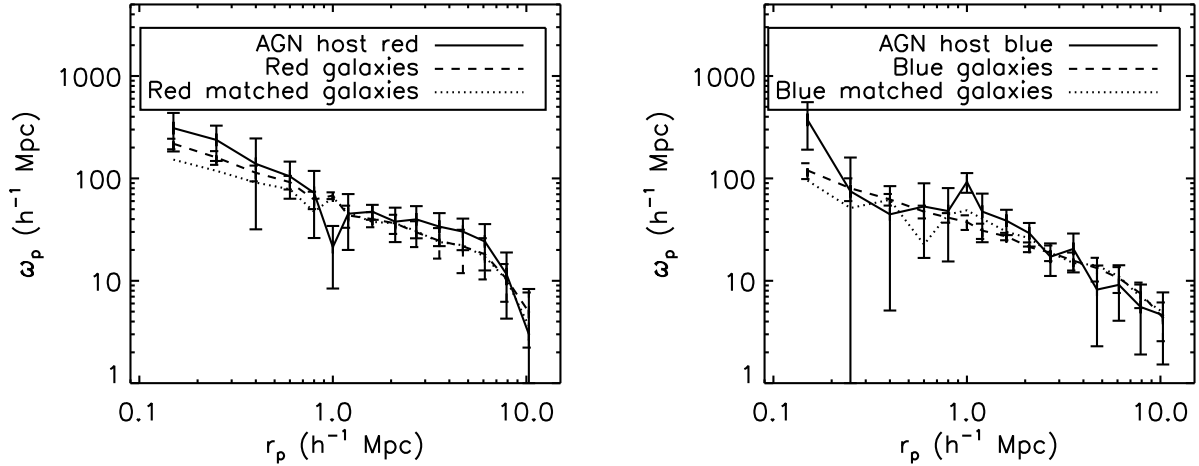}
\caption{The clustering of AGN host galaxies compared to all galaxies of 
a given color.  Left: The projected AGN-galaxy cross-correlation function 
\wprp \ as a function of scale for AGN in red host galaxies (solid line) 
compared to all red galaxies (dashed line) and red galaxies with the same color 
and magnitude distribution (dotted line) at $0.7<z<1.4$.  Right: The 
projected AGN-galaxy cross-correlation for AGN in blue host galaxies (solid line) 
compared to all blue galaxies (dashed line) and blue galaxies with the same
color and magnitude distribution (dotted line) at $0.7<z<1.4$.  Galaxies which 
host AGN are found to be somewhat more clustered than galaxies of the 
same color and magnitude, especially on scales $r_P<1$ \mpch.
}
\label{red_blue_gals}
\end{figure*}

\subsection{Clustering of AGN Compared to Galaxies}

In Figure \ref{gals} we compare the main AGN sample with the clustering 
of red and blue DEEP2 galaxies in the same volume.  Here we have computed
the cross-correlation of red or blue DEEP2 galaxies with all DEEP2 galaxies,
similar to the AGN-galaxy cross-correlation measurement.  We use all red
and blue DEEP2 galaxies, without an absolute magnitude limit, to ensure that
the redshift distribution of the red and blue galaxies is similar to that of 
the X-ray AGN.  The median absolute magnitudes of
the galaxy samples are $M_B=-20.1$ for the blue galaxies and $M_B=-20.7$ 
for the red galaxies.
If we do impose a limit of $M_B<-20$ for the blue galaxies it does not 
change the result; the relative bias is within 1$\sigma$ of
the value using all blue DEEP2 galaxies.  

The X-ray AGN are seen to have a similar clustering amplitude as red DEEP2
galaxies and are more clustered than blue DEEP2 galaxies. The relative bias
of the main AGN sample to red galaxies is $b_{rel}=0.94 \pm0.08$ on large 
scales and $b_{rel}=0.97 \pm0.07$ on all scales, while the relative bias 
to blue galaxies is $b_{rel}=1.48 \pm0.12$ on large scales and 
$b_{rel}=1.61 \pm0.11$ on all scales.  The X-ray AGN are significantly
more clustered than blue galaxies at $z\sim1$.  

The slope of the X-ray AGN correlation function is more similar to that of
blue DEEP2 galaxies than red DEEP2 galaxies \citep{Coil08}, but within 2$\sigma$ 
it is consistent with either.  As seen in Figure \ref{main}, the X-ray AGN 
correlation function may have a steeper slope on small scales, $r_p<0.5$ \mpch, 
compared to all galaxies, but the error bars on the X-ray AGN \wprp \ 
are too large to detect a significant difference.

While the main AGN sample has the same clustering amplitude as red 
galaxies at $z\sim1$, only half (42\%) of X-ray AGN are in red host galaxies.  
This implies that AGN in red galaxies may be more clustered than red galaxies
as a whole and/or that AGN in blue galaxies may be more clustered than
blue galaxies as a whole.
It also implies that galaxies of a given color and luminosity 
distribution that host X-ray AGN may be more clustered than those without AGN.

We can test this by comparing the clustering of the main AGN sample with that 
of DEEP2 galaxies with the same color and magnitude distribution as the AGN 
host sample; the results are shown 
in Figure \ref{all_gals}.  To create the `matched galaxy sample' 
to compare with the main AGN sample, we first measure the joint color-magnitude
distribution of the main AGN sample in $(U-B)$ and $M_B$ space, in bins of 
0.1 in $(U-B)$ color and 0.3 in $M_B$ magnitude.  We then normalize this 
distribution to have a maximum value of 1, making it effectively a relative 
weight map in color-magnitude space.  
We then select all DEEP2 galaxies in the same volume as the main AGN sample,
and for each bin in color-magnitude space we randomly sample the galaxy 
population (without replacement) such that the fraction of the final sample 
in that bin is equal to the weight in the main AGN sample.  
We have tested that creating a matched sample with replacement from the
parent population does not change the results within the 1$\sigma$ error bars.
The matched sample therefore has a nearly
identical joint distribution in color-magnitude space as the main AGN sample.   
The resulting cross-correlation of this `matched galaxy
sample' with all DEEP2 galaxies in the same volume is shown as the dashed 
line in Figure \ref{all_gals}.  The AGN are seen to cluster more
than the matched galaxy sample at the 2.8$\sigma$ level when measured
on scales $0.1<r_p<8$ \mpch, with a relative bias of $b_{rel}=1.28 \pm0.10$, and 
$b_{rel}=1.15 \pm0.09$ on scales $1<r_p<8$ \mpch.  Galaxies of a
given optical color and luminosity which host an AGN are therefore
more clustered than their quiescent counterparts.

The higher clustering amplitude found for AGN in red galaxies compared to 
AGN in blue galaxies could possibly be more fundamentally tied to the 
stellar mass of the host galaxy rather than its optical color; 
determining which is more fundamental is beyond the scope of this paper.  
However, the enhanced clustering amplitude found here for AGN compared to galaxies
with the same color and magnitude distribution can {\it not} be due to a 
difference in stellar mass between the populations; ie, the clustering
difference can not be due to AGN being hosted in more massive galaxies.
The stellar mass of DEEP2 galaxies is tightly correlated with the $(U-B)$ 
color and $M_B$ magnitude (see Figure 10 in \cite{Weiner09}), such that the
matched galaxy samples have similar stellar mass distributions as the
AGN.  

We further compare the clustering of the AGN in red and blue  
galaxies to the full red and blue galaxy samples and also to red and 
blue galaxy samples with the same color and magnitude distribution (`matched' red and 
blue samples)  in Figure \ref{red_blue_gals}.  
The relative biases for these samples are given in Table 2.
There are only minor differences between the biases for 
matched galaxy samples and the full red or blue galaxy samples.  
We find that AGN in red host galaxies are more clustered than both all red 
galaxies and matched red galaxies at the 2.2$\sigma$ level when combining 
measurements from all scales, 
and on large scales they are not significantly more clustered.  
The relative bias is 
similar to that from comparing all AGN to all matched galaxies, but the error bars 
are larger on the red subsamples and the significance is lower. 
Comparing the clustering of AGN in blue host galaxies with matched blue 
galaxies, we do not find a significant difference in their clustering 
properties, though again the relative bias is consistent with that for all AGN
and all matched galaxies; the error bars for the blue subsamples are 
simply too large to say if there is a significant difference in clustering or not.  
The difference that we find between the clustering of the main
AGN sample and all matched galaxies may therefore either be dominated by a 
difference in the clustering of red galaxies with and without AGN or be due
to both red and blue galaxies.

%_______________________________________________________________________

\section{Discussion}

We find here that non-quasar X-ray AGN, taken as an ensemble, have a similar
clustering amplitude as red, quiescent galaxies at $z\sim1$.  
Our results confirm and extend the findings of \cite{Georgakakis07}, who found that 
a smaller sample of X-ray AGN in AEGIS had environments similar to those of
red galaxies.  Not all of the X-ray 
AGN are hosted by red galaxies, however; roughly half are in galaxies whose 
optical color is blue.  Visual inspection of ACS imaging in the AEGIS field 
\citep{Davis07} confirms that the optical light in the X-ray AGN host galaxies 
is not dominated by the AGN itself, but by galaxy light (see also Figure 2 of 
\cite{Nandra07}); the exception to this 
is the 14\% of X-ray AGN that we define as quasars, where the optical 
light is from the AGN itself and is very bright and blue.  We find that 
the X-ray AGN in red host galaxies are more clustered than those in blue
host galaxies, and that when compared to galaxy samples with the same color
and luminosity distribution (and hence, stellar mass), 
those galaxies with X-ray AGN are somewhat more clustered.

\subsection{Comparison to Other X-ray Clustering Studies}

\begin{figure}[t]
\epsscale{1.1}
\plotone{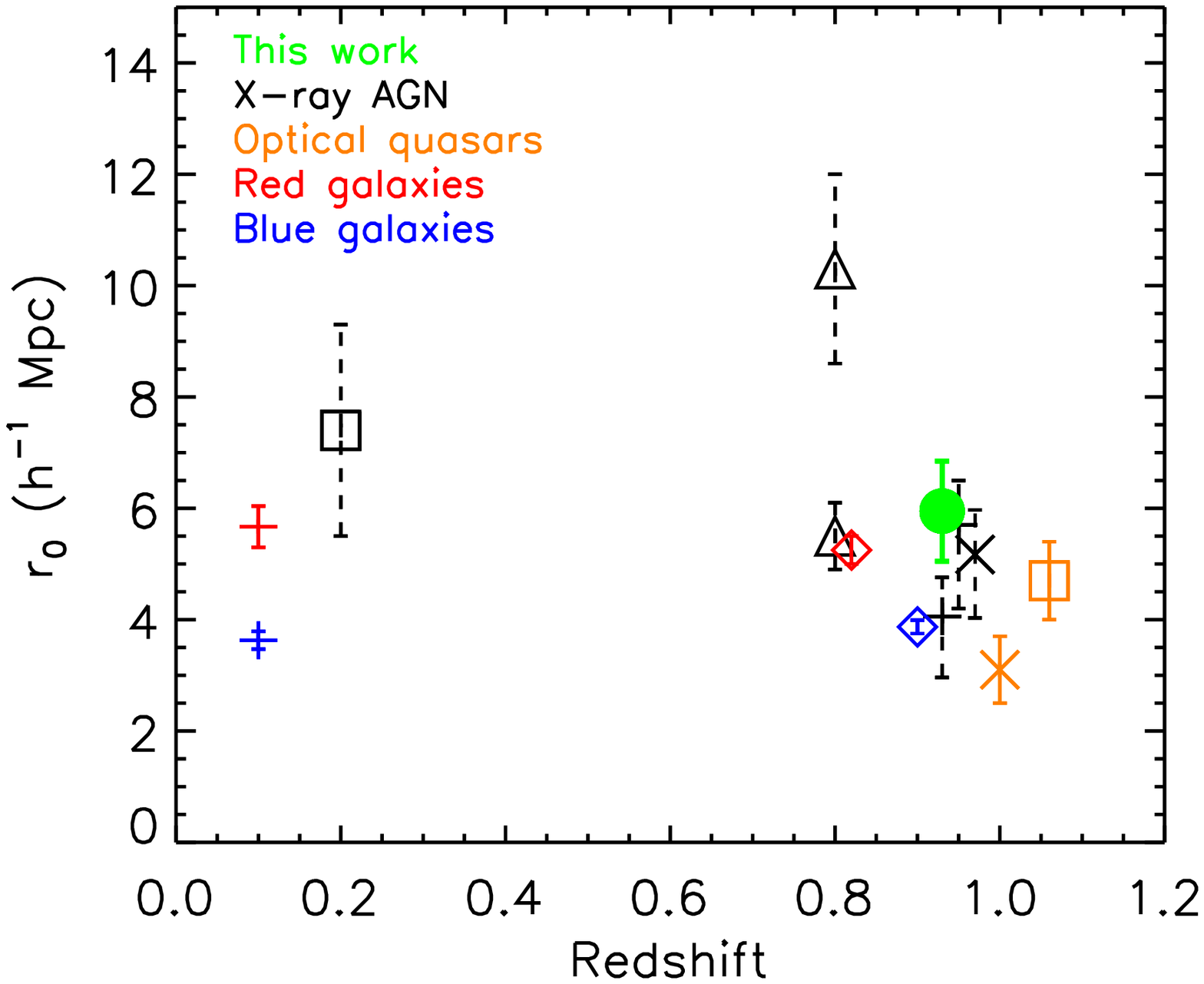}
\caption{Clustering results for various AGN and galaxy surveys with $0<z<1$.  The
comoving correlation
scale length, $r_0$, is plotted for different samples as a function of redshift.
Shown in black with dashed line error bars are  X-ray AGN clustering 
results 
from \cite{Mullis04} (open box), \cite{Gilli05} (triangle), \cite{Yang06} 
(plus sign), and \cite{Gilli09} (X sign).  Results for X-ray AGN from 
this work are shown with a green filled circle.  Results for red and blue galaxies
are shown as red and blue markers for SDSS galaxies from \cite{Zehavi05} (plus sign) and
DEEP2 galaxies from \cite{Coil08} (open diamond).  At $z=0.1$ galaxy clustering 
results from 2dF are similar to those shown for SDSS.  Orange markers indicate 
measures of optically-bright quasar clustering from \cite{Porciani04} 
(orange open box) and 
\cite{Coil07} (orange X sign).  The clustering of X-ray AGN found in this paper 
show that non-quasar X-ray AGN cluster similarly to red galaxies at $z\sim1$ and are
more clustered than blue galaxies.  
}
\label{compare}
\end{figure}

Figure \ref{compare} compares the correlation scale length found for various AGN
and galaxy samples at $z\sim0-1$.  We find here that X-ray AGN at a median 
$z=0.94$ with a 
median luminosity of log $L_{\rm x}=42.8$ erg $s^{-1}$ 
have a correlation length of $r_0=5.95 \pm0.90$ \mpch \ and 
$\gamma=1.66 \pm0.22$.  This clustering scale length is similar to that found by
\cite{Gilli05}, \cite{Yang06}, and \cite{Gilli09} for the CDFN and 
CLASXS and COSMOS fields and differs from that found by \cite{Gilli05} for the
CDFS field; these are the only 
other studies at $z\sim1$ that use spectroscopic redshifts to measure the
correlation function of X-ray AGN.  
At lower redshift, \cite{Mullis04} find 
a consistent correlation length of $r_0=7.4 \pm1.9$ \mpch \ for a fixed slope of 
$\gamma=1.8$; however, a comparison with this result is complicated due to the 
brighter nature of the 
sources (median log $L_{\rm x}=44$ erg $s^{-1}$) and larger measurement scales 
($r_p=5-100$ \mpch).  The correlation slope that we find, $\gamma=1.66 \pm0.22$, 
is consistent with that found by \cite{Gilli05}, \cite{Yang06}, and 
\cite{Gilli09}, which 
are measured on similar scales to those used here.

Our results are in modest disagreement %(at the 2$\sigma$ level) 
with those of \cite{MonteroDorta08}, who 
study the environment-dependence within the red 
sequence of various AGN classes in DEEP2, including optically-selected LINERs, 
Seyferts, and X-ray AGN, and find that while LINERs in red host
galaxies are more likely to be in overdense regions, Seyferts and X-ray
AGN do not have any environment-dependence relative to red galaxies with
the same magnitude distribution as the AGN population.  
In contrast, we find here that X-ray AGN in red galaxies are biased
relative to a galaxy population matched in color and magnitude at 
the 2$\sigma$ level. However, only 36 X-ray sources were used in 
that study, which may account for the discrepancy with our findings here. 

Measuring the cross-correlation function of X-ray sources with
galaxies, as is done here, has several benefits.  One is that the
spatial selection function and redshift completeness for one of the
samples, either the AGN or the galaxies, does not need to be known, as
long as the selection function of the other sample is well characterized.  
By using the cross-correlation of the X-ray sources with galaxies, we do not need 
to include the selection function for the X-ray AGN.  When measuring the
auto-correlation function of X-ray sources, one must carefully model
both the spatial selection function, including the varying X-ray
sensitivity across the field, and the spectroscopic completeness,
 so as not to bias the results. All auto-correlation function measurements
 from {\it Chandra, XMM,} and {\it ROSAT} are potentially affected by the 
sensitivity varying across the field and must carefully address this systematic effect. 

The other benefit of using the cross-correlation function with galaxies in 
the same field is one can sensitively probe the relative bias between AGN
and galaxies as a function of their properties, 
and to first order cosmic variance will cancel in the relative bias.
Cross-correlation functions provide accurate relative clustering measurements and 
do not introduce systematic effects; as shown in \cite{Coil08}, 
the relative bias inferred from cross-correlation measurements is 
consistent with that derived from the ratio of the auto-correlation functions.
Recently, \cite{Hickox09} measured the cross-correlation function of X-ray
AGN with galaxies in the AGES survey (ref) at $z=0.5$ to infer an 
absolute bias of $b=1.34 \pm0.16$ and a mean dark matter halo mass of $\sim1 
\times 10^{13} h^{-1} M_{\sun}$, similar to the results found here.

Several studies have investigated the dependence of X-ray AGN clustering
on the hardness or hardness ratio of the source.
Within the errors, we find no significant difference between AGN samples 
divided according to hardness ratio.  
\cite{Gilli05} and \cite{Gilli09} also find no significant difference in the 
clustering strength of soft and
hard X-ray sources in the CDFN, CDFS, and COSMOS fields.  
\cite{Yang03} detect a higher 
angular clustering signal for hard sources
compared to soft sources, though without a well-known redshift distribution 
for the hard and soft samples this can not be interpreted as a significant
difference in the intrinsic clustering properties.  If, for example, the
harder sources had a narrower redshift distribution, that would 
account for the different angular clustering signals.  

We also find no significant dependence of clustering amplitude on X-ray 
luminosity within the range probed here.  \cite{Plionis08} claim to find
a trend with brighter X-ray sources having higher clustering amplitudes, 
but these results rely on angular clustering measures without known redshift
distributions, and the authors infer extremely large and implausible 
correlation lengths at $z\sim1$ of $\sim10-20$ \mpch.  
\cite{Yang06} detect a `weak' correlation between clustering
amplitude and X-ray luminosity when comparing their 
X-ray AGN clustering results with that of 2dF quasars \citep{Croom04}, 
where X-ray luminosities for the quasars are estimated from 
their bolometric luminosities.  They find that quasars are somewhat more
clustered than X-ray AGN, though at $<2\sigma$ significance.  
This differs with what we find 
below (and the results shown in Figure \ref{compare}), when comparing the 
clustering of X-ray AGN and quasars in the 
same volume at the same redshift.  The $z=0.2$ results of \cite{Mullis04} 
for brighter X-ray AGN do indicate a slightly higher correlation length 
than those at $z\sim1$, but that may be due to evolution and growth
of structure, not intrinsic luminosity-dependent clustering.

The lack of a convincing relation between X-ray luminosity and clustering 
strength is important for theoretical models, as the X-ray luminosity is 
related to the black hole accretion rate for a fixed black hole mass.
This may indicate then that there is not a strong correlation between accretion
rate and host halo mass, i.e., between how much gas is accreted onto a  
black hole at a given time and how much gas potentially exists in the 
reservoir of the halo.

\subsection{Comparison to Optical AGN Clustering Studies}

A comparison of the clustering properties of different kinds of AGN can lead 
to a clearer understanding of their respective host galaxies, fueling 
mechanisms and AGN lifetimes. 
The clustering of X-ray AGN was not previously known to high enough
precision to make a detailed comparison with other AGN samples.  
Among optically-selected AGN, clustering studies have been performed both for
quasars at intermediate and high redshift and lower-luminosity Seyferts at 
low redshift. 

Using the same techniques as here, \cite{Coil07} find that 
spectroscopic broad-line SDSS quasars in the DEEP2 volume at $0.7<z<1.4$ 
cluster like blue galaxies rather than red galaxies (at 2$\sigma$ significance),
with $r_0=3.1 \pm0.6$ \mpch.  
Comparing the scale length found here for non-quasar X-ray AGN with that 
found by \cite{Coil07} for quasars, the X-ray AGN are more clustered 
than quasars at the 2.6$\sigma$ level.  This comparison should be quite robust,
as the same sample of galaxies from the DEEP2 survey is used as a tracer of
large-scale structure in both analyses, and therefore the survey selection,
and design are identical, as are the scales probed.  
This comparison suggests that 
high-accretion AGN, observed as optically bright quasars, are likely hosted in
star-forming galaxies, while X-ray AGN are hosted both by 
massive blue galaxies and red galaxies, as seen in Figure \ref{cmd} (see also 
\cite{Nandra07}), reflecting a general correlation between the
amount of star formation in a galaxy and the accretion rate of its central
black hole (see e.g., \cite{Kauffmann07} and \cite{Silverman08} for other
observational signatures of this general correlation).  As both quasar fueling 
and star formation require the presence of cold gas, it is reasonable that they should
be found in similar environments.  {\it If} optically-bright,
high-accretion quasars and lower-accretion X-ray AGN
are similar objects at different evolutionary stages, then these results
are consistent with a scenario in which a quasar resides in a star forming
galaxy in the blue cloud before having its star formation quenched and moving
to the red sequence with a lower luminosity X-ray AGN.  The timescale between
the quasar and lower-accretion X-ray AGN stages would have to be on the order 
of several Gyr, however, to allow enough time for the clustering amplitude to 
increase due to gravity.

Other studies of quasars at $z\sim1$  \citep[e.g.,][]{Porciani04,Myers06,
daAngela05} find somewhat higher 
correlation lengths (consistent at the 2$\sigma$ level with the \cite{Coil07} results) 
using the quasar auto-correlation function, though
usually with larger error bars due to the low number density
of quasars.  The most constraining of these results is from \cite{Porciani04}, using
the 2QZ survey, who find $r_0=4.7 \pm0.7$ \mpch \ at $z=1.06$ (shown as an orange
open box in Figure \ref{compare}).  This result is consistent with \cite{Coil07} but 
allows a correlation length that matches either red or blue galaxies, given the error bars.
This is generally true of other quasar clustering results at this redshift.
Quasar clustering studies generally find a correlation slope, $\gamma$, of $\sim1.7$, very 
similar to that of blue galaxies but lower than that of red galaxies at 
$z\sim1$ \citep{Coil08}, which implies that quasars can not predominantly 
be hosted by red galaxies.  

While a comparison of the quasar and AGN clustering results from DEEP2 point to
AGN possibly being more clustered than quasars, a more precise measurement of the
quasar clustering amplitude is needed to more firmly identify the host galaxy types 
and dark matter halo masses of quasars.  
The fundamental limitation in our ability to 
more precisely measure the quasar correlation function is the 
low number density of quasars; this makes cross-correlations of quasar
samples with more abundant galaxy samples particularly effective.  Future 
precise measurements of the quasar correlation function will likely come from
wide-area galaxy redshift surveys at intermediate and high redshift.

\subsection{Black Hole Masses for Quasars and X-ray AGN}

The 2$\sigma$ difference in clustering amplitude measured at $z\sim1$ for 
quasars and X-ray AGN, if real, would reveal a difference in the mean host 
dark matter halo mass of these two populations, which could reflect a 
difference in the mean black hole masses for quasars and X-ray AGN. 
To investigate this we estimate black hole masses for the quasars used 
in \cite{Coil07} and the X-ray AGN used in this study.

To estimate black hole masses for the broad-line SDSS quasar population used in 
\cite{Coil07}, we use properties measured in the SDSS spectra, in particular 
the {\rm Mg II} 2800 \AA \ linewidth and UV flux, which are available for all of
the quasars used in that study.  
The following estimator from \cite{Mclure04} is used, which is calibrated at low 
redshift using reverberation mapping:

\begin{equation}
{\rm \frac{M_{BH}}{M_{\sun}}=3.2 \left(\frac{\lambda L_{3000}}{10^{37}W}\right)^{0.62} \left[\frac{FWHM(Mg II)}{km \ s^{-1}} \right]^2},
\end{equation}

where L$_{3000}$ is the restframe continuum luminosity at 3000 \AA.
We find that the SDSS quasars in the \cite{Coil07} sample have a median black hole 
mass of $M_{BH}=3.0 \times 10^8 M_{\sun}$ and a mean mass of 
$\overline{M}_{BH}=4.3 \times 10^8 M_{\sun}$, with a 1$\sigma$ standard deviation of 
$3.4 \times 10^8 M_{\sun}$, which agrees well with the
masses of SDSS quasars at $z\sim1$ found by \cite{Mclure04}.  
There may be systematic biases associated with this method, particularly given
that we are applying it at a different redshift than it was calibrated, 
but the estimated masses should be within a factor 
of $\sim$2-3 of the true masses.

Black hole mass estimates for the X-ray AGN sample used in this paper 
are presented and discussed in \cite{Bundy08},
where the following relation between black hole mass and bulge infrared luminosity
is used \citep{Graham07}:
\begin{equation}
{\rm log\left(\frac{M_{BH}}{M_{\sun}}\right) = -0.37 (M_{K_{vega}}+24) + 8.29}.
\end{equation}

K-band photometry for the Chandra sources is from a Palomar/WIRC survey of the
DEEP2 fields \citep{Bundy06}.  To estimate the black hole masses we 
assume no evolution in the black hole-bulge mass relation with redshift. 
 We also assume that the host galaxies of the X-ray AGN are 
bulge-dominated and 
that the bulk of the K-band light is from the bulge.  This is a reasonable
assumption, as HST imaging shows that most of the X-ray AGN are in 
bulge-dominated galaxies 
\citep{Pierce07}.  There is a systematic effect, however, that would drive 
the black hole mass estimates lower as less of the K-band light is in the bulge.
Here again we expect these black hole masses to be within a factor of 
$\sim$2-3 of the true masses.  We find that the median black hole mass for 
the X-ray AGN is $M_{BH}=4.7 \times 10^8 M_{\sun}$ and the mean is 
$\overline{M}_{BH}=6.2 \times 10^8 M_{\sun}$, with a 1$\sigma$ standard deviation of 
$5.8 \times 10^8 M_{\sun}$.

The inferred X-ray AGN black hole masses are comparable to the quasar black hole 
masses, which could imply that they are in similar stellar mass galaxies.
It may be surprising then if the non-quasar X-ray AGN are more clustered 
than the quasars.  If true, this could indicate that quasars are more prevalent
in lower mass halos because those halos have more fuel available to power the
black hole.
   
\cite{Ferrarese02} finds a scaling between black hole mass and host
dark matter halo mass at low redshift:
\begin{equation}
{\rm M_{BH}/10^8 M_{\sun} = 0.67 \  (M_{halo} / 10^{12} M_{\sun})^{1.82}},
\end{equation}
where results from \cite{Seljak02} have been used for halo profile fits 
derived from weak lensing measurements.
Given our rough black hole mass estimates above, 
this relation implies that both the SDSS quasars and non-quasar 
X-ray AGN should be in
$\sim3 \times 10^{12} M_{\sun}$ halos, if there is no evolution in this 
relation to $z\sim1$.  As presented in \cite{Coil08}, from clustering
measurements the minimum dark
matter halo mass for blue galaxies at $z\sim1$ is 
$M_{min}\sim5 \times 10^{11} h^{-1} M_{\sun}$ and for red galaxies is 
 $M_{min}\sim2 \times 10^{12} h^{-1} M_{\sun}$.   The corresponding {\it mean} 
halo masses for blue and red galaxies 
are $\overline{M}\sim1.8 \times 10^{12} M_{\sun}$
and $\overline{M}\sim5 \times 10^{13} M_{\sun}$, respectively, for $h=0.7$.  
Assuming that the scaling law between black hole mass and dark matter halo
host mass does not evolve to $z\sim1$, this would imply that quasars and 
non-quasar X-ray AGN reside either primarily in blue galaxies or in a mix
of red and blue galaxies, which is consistent with the location of the
X-ray AGN in the color-magnitude diagram.

\subsection{X-ray AGN and Green Galaxies}

The clustering of `green' galaxies in the minimum of the optical color
bimodality at $z\sim1$ has been 
measured by \cite{Coil08}.  A comparison of the green 
galaxies -- which may be in transition from the blue, star forming 
galaxy population to the red, quiescent galaxy population -- to our 
measurements for the X-ray AGN shows that both populations
have a high clustering amplitude on large scales and therefore similar values
of the correlation length ($r_0=5.17 \pm0.42$ \mpch \ for green
galaxies with $M_B<-20$), comparable to red galaxies.  Unlike red galaxies,
however,  green galaxies have a relatively shallow clustering slope, 
 $\gamma=1.59 \pm0.08$, similar to that for blue, star-forming galaxies.  
The X-ray AGN clustering slope is not well constrained here and is consistent 
with the slope of either red or blue galaxies.  

The location of the X-ray AGN in the optical color-magnitude 
diagram (see Figure \ref{cmd}) indicates that there
is likely to be overlap between the X-ray AGN and green galaxy populations. 
Their clustering properties reinforce this; 
both populations reside in massive halos.  The shallow 
slope of the green galaxy correlation function likely reflects a
greater tendency for green galaxies to be found near the outskirts of 
dark matter halos, rather than their centers, compared to red galaxies.
The same may be true for the X-ray AGN, though the apparent rise of \wprp \ 
on small scales 
might indicate that X-ray AGN are actually more likely to be central galaxies.
However, given the large uncertainties in \wprp \ in the innermost bins we can
 not distinguish between these scenarios here.

\subsection{X-ray AGN in Groups}

The higher clustering amplitude, found here at a significance of 2.8 $\sigma$,
for X-ray AGN compared to 
galaxies of the same color and magnitude distribution suggests that a higher
fraction of X-ray AGN host galaxies are in groups than all galaxies 
of the same color and magnitude, as galaxies in groups have a higher clustering
amplitude \citep{Coil05, Yang05a}.  As discussed in Section 5.3, it can not be
due to a difference in the stellar mass distribution of the AGN host galaxies
compared to the matched galaxy sample, as they have similar stellar masses (as
inferred from their host galaxy colors and luminosities). 
 For a given massive galaxy sample, 
whether star forming or quiescent, it is the galaxies that reside in more 
massive halos, and therefore are more likely to be in groups, that host 
X-ray AGN.  

\cite{georgakakis08} measure the fraction of X-ray AGN at 
$z\sim1$ in groups in AEGIS using the same {\it Chandra} and DEEP2 galaxy 
data used here.  Using a DEEP2 group catalog \citep{Gerke05} that 
has been tested and calibrated with simulations and mock galaxy catalogs, 
they find that when compared to the full DEEP2 galaxy population X-ray AGN 
are preferentially found in groups, but when compared to galaxies of the same
color and luminosity distribution of the AGN hosts this difference drops to 
the $<2\sigma$ level.  This is almost certainly consistent with what we find here, 
 as the higher clustering amplitude found for X-ray AGN host galaxies
when compared to all galaxies of the same color and magnitude distribution
  does not imply that {\it all} X-ray AGN are in groups identified by
\cite{Gerke05}, but rather that they generally reside in higher mass host halos which 
have a stronger likelihood of being in a group.  Additionally, random errors are 
larger in the study of \cite{georgakakis08}, due to small number statistics for the group
catalog, such that what is significantly detected here may be seen at lower significance 
in the group study.

\subsection{The Fueling of X-ray AGN}

We find that non-QSO X-ray AGN host galaxies reside in more massive dark matter 
halos than the full population of galaxies of the same color and magnitude
distribution.  This likely reflects a higher incidence of X-ray AGN in 
galaxy groups than in isolated galaxies.  
Residing in a galaxy group may lead to accretion onto a 
central black hole, if galaxy mergers (which are more likely
to occur in groups) generally supply gas to the central AGN
\citep[e.g.,][]{Barnes92, Mihos96, Springel05c}. 
Alternatively, as X-ray AGN favor host dark matter halos that are on the 
massive end of the distribution for given observed galaxy properties, 
their presence in these halos may not reflect interactions with the environment,
but rather a tendency for more massive black holes in otherwise similar galaxies
to be found in more massive halos, possibly due to the presence of more gas 
that could fuel the black hole.  For X-ray AGN in red host 
galaxies, there may be hot gas on large scales which could cool on small 
scales near the black hole to provide fuel.

The host galaxy morphologies of $\sim$60 X-ray AGN in AEGIS is studied by 
\cite{Pierce07} (see also Georgakakis et al. in preparation), who find that 
roughly half of the host galaxies are elliptical
or bulge-dominated, and that while the fraction that are classified as mergers 
is higher than for the field (at the 2$\sigma$ level), this accounts for only
$\sim$20\% of the population.  It is therefore not clear that X-ray AGN are 
necessarily fueled by on-going or recent major mergers; they 
may be fueled 
by minor mergers or, in the case of those in red host galaxies, they may be 
fueled by small amounts of hot gas that remain long after a major merger.
Indeed, roughly half of the X-ray AGN lie in red host galaxies that may have 
been on the red sequence for a long time and possibly not have undergone a major 
merger for many Gyr.  To reconcile our results with a scenario in which X-ray AGN are 
fueled primarily by major mergers, one would require that those in red 
host galaxies had recently migrated to the red sequence 
\citep[e.g.,][]{Salim07,Schawinski07, Shen07, Georgakakis08b, Kocevski08}. 

As discussed above, the results presented here, when combined with those 
of \cite{Coil07}, 
are consistent with an evolutionary sequence in which bright quasars 
are ignited while being hosted by star forming galaxies and later 
evolve to be lower-luminosity AGN, detected in X-rays but not as broad-line 
optical sources, hosted by massive galaxies in either
the bright or red end of the blue cloud or on the red sequence.  
This evolutionary sequence is not a unique explanation of the observations, 
i.e., X-ray AGN may be a different population than quasar relics, 
but if some fraction of blue star-forming
galaxies have their star formation quenched and migrate to the red sequence, 
then it is plausible that some quasars may eventually become lower-luminosity
AGN detected not by their optical light but by their X-ray emission.  In 
this picture the AGN type could be due to gas availability in the host galaxy,
which may be related to host halo mass \citep[e.g.,][]{Croton06,Dekel06}.

The results presented here and in \cite{Coil07} are generally 
consistent with the cartoon model shown in \cite{Hickox09} for how 
galaxies and AGN evolve with time.  In this cartoon model 
optically-bright quasars are hosted by on-going disk galaxy mergers 
and immediately precede an optically-faint X-ray AGN phase, which 
evolves into an early-type galaxy.

\subsection{Comparison to Theoretical Models of AGN Evolution}

Theoretical models of AGN formation and evolution can yield 
measurably different predictions for the local 
environments and clustering properties of AGN. 
\cite{Kauffmann02} use a semi-analytic model in
which AGN are fueled by galaxy mergers, where the peak AGN
luminosity depends on the mass of gas accreted by the black hole,
which in turn depends on the host halo mass.
This leads to a natural prediction that brighter AGN reside in
more massive halos, such that the AGN
clustering amplitude should be strongly luminosity-dependent.  This
model is not supported by observations, which generally show 
a lack of a strong correlation between AGN clustering amplitude and
luminosity, except at the very bright end
\citep[e.g.,][]{Croom02,Porciani06,Myers07,Shen08}.

\cite{Hopkins05a} present an alternative model
in which bright and faint AGN are in similar physical
systems but are in different stages of their life cycles.  
This model predicts that faint and bright AGN 
should reside in similar--mass dark matter halos and that quasar clustering
should depend only weakly on luminosity \citep{Lidz06}.  That general 
prediction agrees well qualitatively with observations of quasar clustering, 
but the model also predicts that lower luminosity AGN should 
have an equal or lower clustering amplitude than bright AGN, 
which is not well-supported when comparing our results here for 
non-quasar X-ray AGN with the results for quasars in \cite{Coil07}.
In the specific model presented by \cite{Hopkins08}, AGN are detected
in X-rays while obscured by dust just after a major merger, 
before undergoing an optically-bright quasar phase a short time later;
this picture is also not well-supported by our results.

The semi-analytic model of \cite{Croton06} combines a 
 prescription of merger-driven black hole growth similar to 
\cite{Kauffmann00}
with an independent mode for hot gas accretion in large halos which
accounts for the fueling of lower-luminosity AGN in massive halos.
This model assumes that some fraction of cold gas 
must be present to trigger a bright quasar phase during a galaxy merger;
the black hole itself must also be massive such that the luminosity remains
sub-Eddington.  Quasars, in this model, can be strongly clustered at high 
redshift when cold gas fractions are presumably high, but as their host dark
matter halos grow cooling becomes more inefficient as the virial temperature of
the halo increases, and the cold gas supply is suppressed above a given threshold
mass.  Quasars will only be found in halos below the threshold mass, and 
thus relatively gas-free red galaxies are not expected to shine as 
quasars; only those mergers that occur in lower mass halos, presumably outside of group 
environments at $z<1$,  will contain sufficient cold gas to fuel a quasar.  
This model therefore predicts that quasars should cluster 
similarly to massive star-forming galaxies at a given redshift, where both 
the star formation and quasar activity are fueled by cold gas, while massive
halos host lower-luminosity AGN that are fueled (possibly relatively 
inefficiently) by hot gas (which may cool on small scales near the black hole).  X-ray AGN, in this picture, therefore should cluster like quasars did
at an earlier epoch.
There are no quantitative predictions from this model for the clustering
of, say, the full X-ray AGN population versus bright quasars, but the model 
is qualitatively consistent with our results and the relatively higher clustering
seen for quasars at high redshift \citep{Myers07,Shen07,Croom05}.

\cite{Thacker08} present a model for quasar fueling and feedback that 
quantitatively matches observations of quasar and X-ray clustering 
reasonably well.  
In their model bright AGN (both quasars and bright X-ray AGN) are the 
result of mergers and are augmented by feedback from AGN outflows.  They do 
not present predictions for $z<1.2$, but their model matches well the observed
clustering of quasars at $z=1.5-2$.  Their prediction for the correlation
length of X-ray AGN at $z=1.75$ with a luminosity 
$L_{\rm X}=3.2 \times 10^{43} erg s^{-1}$ is 3.6 \mpch, significantly lower 
than what we find here at $z=0.9$.  However, as it is not clear that 
the bulk of the X-ray AGN studied here are the result of major mergers, this
model may not be entirely applicable to the results presented here.

%_______________________________________________________________________

\section{Conclusions}

Using the cross-correlation of galaxies and {\it Chandra-}detected AGN in the
AEGIS survey, we are able to determine the clustering of
X-ray AGN at $z\sim1$ 
much more precisely than studies that measure the clustering of AGN using 
the auto-correlation function of AGN alone.  We 
find that non-quasar X-ray AGN with a median log $L_{\rm X}\sim43$ erg $s^{-1}$ 
have a clustering scale length of 
$r_0=5.95 \pm0.90$ \mpch \ and slope of $\gamma=1.66 \pm0.22$,
with a bias that corresponds to a minimum dark matter halo mass of
$M_{min}=5 \ (+5/-3) \times 10^{12} h^{-1} M_{\sun}$.
The clustering amplitude of X-ray AGN is consistent with that 
of red and `green' galaxies at the same redshift and is greater 
than that of blue galaxies, 
even though roughly half of the X-ray AGN are hosted by blue, star-forming 
galaxies.  We do not find a significant difference in clustering amplitude within the
AGN population when divided into subsamples based on on optical luminosity, 
X-ray luminosity, or hardness ratio, within the ranges probed here.  

The clustering strength of X-ray AGN is primarily determined by the host
galaxy color; there is a smaller but significant effect due to a 
difference in clustering amplitude for galaxies
that host X-ray AGN compared to matched samples of galaxies with 
the same color and luminosity
distribution.  This implies that galaxies that host X-ray AGN are in more
massive dark matter halos than those that do not, for fixed host galaxy properties;
i.e., galaxies that host X-ray AGN are more likely to be in groups when
compared to galaxies of the same color and luminosity distribution.

By comparing to previous results on the clustering of 
optically-selected quasars in the DEEP2 fields, we find that non-quasar 
X-ray AGN are more clustered than optically-selected quasars at the same
redshift at the $2.6\sigma$ level.  
Our results are consistent with quasars occurring in blue star-forming galaxies
which later settle onto the red sequence and potentially host lower-luminosity
X-ray AGN,
{\it if} they are similar objects at different
evolutionary stages.  While clustering measures of different types of AGN
can be used to constrain their host galaxy types, halo masses, and 
fueling mechanisms, precise measurements await larger galaxy and AGN redshift
surveys at intermediate and high redshift.

%_______________________________________________________________________

\acknowledgements

ALC would like to thank David Schlegel for measuring spectral 
quantities used to determine black hole masses for the SDSS quasars, 
Kevin Bundy for providing black hole mass estimates for the 
X-ray AGN, and Richard Cool for his help with the Hectospec data reduction.  
ALC would also like to thank Aaron Barth, Daniel Eisenstein, 
Jenny Greene, and Juna Kollmeier for useful discussions.

\

ALC was supported by NASA through Hubble Fellowship grant
HF-01182.01-A, awarded by the Space Telescope Science Institute, which
is operated by the Association of Universities for Research in
Astronomy, Inc., for NASA, under contract NAS 5-26555.  The DEEP2 team
acknowledges support from NSF grants AST-0507483 and AST-0808133 and
{\it Chandra} NASA grant GO8-9129A.  The DEIMOS spectrograph was
funded by a grant from CARA (Keck Observatory), an NSF Facilities and
Infrastructure grant (AST92-2540), the Center for Particle
Astrophysics and by gifts from Sun Microsystems and the Quantum
Corporation.  Much of the data presented herein were obtained at the
W.M. Keck Observatory, which is operated as a scientific partnership
among the California Institute of Technology, the University of
California and the National Aeronautics and Space Administration. The
Observatory was made possible by the generous financial support of the
W.M. Keck Foundation.  Additional observations reported here were
obtained at the MMT Observatory, a joint facility of the University of
Arizona and the Smithsonian Institution.


\begin{thebibliography}{}

\bibitem[{Akylas}, {Georgantopoulos}, \& {Plionis}(2000){Akylas},
  {Georgantopoulos}, and {Plionis}]{Akylas00}
{Akylas}, A., {Georgantopoulos}, I., \& {Plionis}, M. 2000, \mnras, 318, 1036

\bibitem[{Alexander} {et~al.}(2001){Alexander} {\em et~al.}]{Alexander01}
{Alexander}, D.~M., et~al. 2001, \aj, 122, 2156

\bibitem[{Barger} {et~al.}(2003){Barger} {\em et~al.}]{Barger03}
{Barger}, A.~J., et~al. 2003, \aj, 126, 632

\bibitem[{Barnes} \& {Hernquist}(1992){Barnes} and {Hernquist}]{Barnes92}
{Barnes}, J.~E., \& {Hernquist}, L. 1992, \araa, 30, 705

\bibitem[{Basilakos} {et~al.}(2004){Basilakos}, {Georgakakis}, {Plionis}, and
  {Georgantopoulos}]{Basilakos04}
{Basilakos}, S., {Georgakakis}, A., {Plionis}, M., \& {Georgantopoulos}, I.
  2004, \apjl, 607, L79

\bibitem[{Basilakos} {et~al.}(2005){Basilakos}, {Plionis}, {Georgakakis}, and
  {Georgantopoulos}]{Basilakos05}
{Basilakos}, S., {Plionis}, M., {Georgakakis}, A., \& {Georgantopoulos}, I.
  2005, \mnras, 356, 183

\bibitem[{Brandt} \& {Hasinger}(2005){Brandt} and {Hasinger}]{Brandt05}
{Brandt}, W.~N., \& {Hasinger}, G. 2005, \araa, 43, 827

\bibitem[{Bundy} {et~al.}(2006){Bundy} {\em et~al.}]{Bundy06}
{Bundy}, K., et~al. 2006, \apj, 651, 120

\bibitem[{Bundy} {et~al.}(2008){Bundy} {\em et~al.}]{Bundy08}
{Bundy}, K., et~al. 2008, \apj, 681, 931

\bibitem[{Cattaneo} {et~al.}(2006){Cattaneo} {\em et~al.}]{Cattaneo06}
{Cattaneo}, A., et~al. 2006, \mnras, 370, 1651

\bibitem[{Coil} {et~al.}(2004a){Coil} {\em et~al.}]{Coil04}
{Coil}, A.~L., et~al. 2004a, \apj, 617, 765

\bibitem[{Coil} {et~al.}(2004b){Coil} {\em et~al.}]{Coil04xisp}
{Coil}, A.~L., et~al. 2004b, \apj, 609, 525

\bibitem[{Coil} {et~al.}(2006a){Coil} {\em et~al.}]{Coil06lum}
{Coil}, A.~L., et~al. 2006a, \apj, 644, 671

\bibitem[{Coil} {et~al.}(2006b){Coil} {\em et~al.}]{Coil05}
{Coil}, A.~L., et~al. 2006b, \apj, 638, 668

\bibitem[{Coil} {et~al.}(2007){Coil} {\em et~al.}]{Coil07}
{Coil}, A.~L., et~al. 2007, \apj, 654, 115

\bibitem[{Coil} {et~al.}(2008){Coil} {\em et~al.}]{Coil08}
{Coil}, A.~L., et~al. 2008, \apj, 672, 153

\bibitem[{Comastri} \& {Fiore}(2004){Comastri} and {Fiore}]{Comastri04}
{Comastri}, A., \& {Fiore}, F. 2004, \apss, 294, 63

\bibitem[{Croom} {et~al.}(2004){Croom} {\em et~al.}]{Croom04}
{Croom}, S., et~al. 2004, In ASP Conf. Ser. 311: AGN Physics with the Sloan
  Digital Sky Survey, p. 457

\bibitem[{Croom} {et~al.}(2002){Croom} {\em et~al.}]{Croom02}
{Croom}, S.~M., et~al. 2002, \mnras, 335, 459

\bibitem[{Croom} {et~al.}(2005){Croom} {\em et~al.}]{Croom05}
{Croom}, S.~M., et~al. 2005, \mnras, 356, 415

\bibitem[{Croton} {et~al.}(2006){Croton} {\em et~al.}]{Croton06}
{Croton}, D.~J., et~al. 2006, \mnras, 365, 11

\bibitem[{da {\^A}ngela} {et~al.}(2005){da {\^A}ngela} {\em
  et~al.}]{daAngela05}
{da {\^A}ngela}, J., et~al. 2005, \mnras, 360, 1040

\bibitem[{Davis} {et~al.}(2003){Davis} {\em et~al.}]{Davis03}
{Davis}, M., et~al. 2003, Proc. SPIE, 4834, 161

\bibitem[{Davis} {et~al.}(2007){Davis} {\em et~al.}]{Davis07}
{Davis}, M., et~al. 2007, \apjl, 660, L1

\bibitem[{Davis} \& {Peebles}(1983){Davis} and {Peebles}]{Davis83}
{Davis}, M., \& {Peebles}, P.~J.~E. 1983, \apj, 267, 465

\bibitem[{Davis} {et~al.}(2000){Davis}, {Newman}, {Faber}, and
  {Phillips}]{Davis00}
{Davis}, M., {Newman}, J.~A., {Faber}, S.~M., \& {Phillips}, A.~C. 2000, In
  Proceedings of the ESO/ECF/STSCI Workshop on Deep Fields, Garching (Publ:
  Springer)

\bibitem[{Dekel} \& {Birnboim}(2006){Dekel} and {Birnboim}]{Dekel06}
{Dekel}, A., \& {Birnboim}, Y. 2006, \mnras, 368, 2

\bibitem[{Dekel} \& {Birnboim}(2008){Dekel} and {Birnboim}]{Dekel08}
{Dekel}, A., \& {Birnboim}, Y. 2008, \mnras, 383, 119

\bibitem[{Donley} {et~al.}(2005){Donley} {\em et~al.}]{Donley05}
{Donley}, J.~L., et~al. 2005, \apj, 634, 169

\bibitem[{Faber} {et~al.}(2003){Faber} {\em et~al.}]{Faber03}
{Faber}, S., et~al. 2003, Proc. SPIE, 4841, 1657

\bibitem[{Ferrarese}(2002){Ferrarese}]{Ferrarese02}
{Ferrarese}, L. 2002, \apj, 578, 90

\bibitem[{Ferrarese} \& {Ford}(2005){Ferrarese} and {Ford}]{Ferrarese05}
{Ferrarese}, L., \& {Ford}, H. 2005, Space Science Reviews, 116, 523

\bibitem[{Ferrarese} \& {Merritt}(2000){Ferrarese} and {Merritt}]{Ferrarese00}
{Ferrarese}, L., \& {Merritt}, D. 2000, \apjl, 539, L9

\bibitem[{Gebhardt} {et~al.}(2000){Gebhardt} {\em et~al.}]{Gebhardt00}
{Gebhardt}, K., et~al. 2000, \apjl, 539, L13

\bibitem[{Georgakakis} {et~al.}(2007){Georgakakis} {\em et~al.}]{Georgakakis07}
{Georgakakis}, A., et~al. 2007, \apjl, 660, L15

\bibitem[{Georgakakis} {et~al.}(2008a){Georgakakis} {\em
  et~al.}]{Georgakakis08b}
{Georgakakis}, A., et~al. 2008a, \mnras, 385, 2049

\bibitem[{Georgakakis} {et~al.}(2008b){Georgakakis} {\em
  et~al.}]{georgakakis08}
{Georgakakis}, A., et~al. 2008b, \mnras, p. 1197

\bibitem[{Gerke} {et~al.}(2005){Gerke} {\em et~al.}]{Gerke05}
{Gerke}, B.~F., et~al. 2005, \apj, 625, 6

\bibitem[{Gilli} {et~al.}(2005){Gilli} {\em et~al.}]{Gilli05}
{Gilli}, R., et~al. 2005, \aap, 430, 811

\bibitem[{Gilli} {et~al.}(2009){Gilli} {\em et~al.}]{Gilli09}
{Gilli}, R., et~al. 2009, Accepted to A\&A

\bibitem[{Gilli}, {Comastri}, \& {Hasinger}(2007){Gilli}, {Comastri}, and
  {Hasinger}]{Gilli07}
{Gilli}, R., {Comastri}, A., \& {Hasinger}, G. 2007, \aap, 463, 79

\bibitem[{Graham}(2007){Graham}]{Graham07}
{Graham}, A.~W. 2007, \mnras, 379, 711

\bibitem[{Heckman} {et~al.}(2005){Heckman} {\em et~al.}]{Heckman05}
{Heckman}, T.~M., et~al. 2005, \apj, 634, 161

\bibitem[{Hickox} {et~al.}(2009){Hickox} {\em et~al.}]{Hickox09}
{Hickox}, R.~C., et~al. 2009, \apj, 696, 891

\bibitem[{Hopkins} {et~al.}(2005){Hopkins} {\em et~al.}]{Hopkins05a}
{Hopkins}, P.~F., et~al. 2005, \apjl, 625, L71

\bibitem[{Hopkins} \& {Hernquist}(2006){Hopkins} and {Hernquist}]{Hopkins06}
{Hopkins}, P.~F., \& {Hernquist}, L. 2006, \apjs, 166, 1

\bibitem[{Hopkins} {et~al.}(2008){Hopkins}, {Hernquist}, {Cox}, and {Kere{\v
  s}}]{Hopkins08}
{Hopkins}, P.~F., {Hernquist}, L., {Cox}, T.~J., \& {Kere{\v s}}, D. 2008,
  \apjs, 175, 356

\bibitem[{Kauffmann} {et~al.}(2007){Kauffmann} {\em et~al.}]{Kauffmann07}
{Kauffmann}, G., et~al. 2007, \apjs, 173, 357

\bibitem[{Kauffmann} \& {Haehnelt}(2000){Kauffmann} and
  {Haehnelt}]{Kauffmann00}
{Kauffmann}, G., \& {Haehnelt}, M. 2000, \mnras, 311, 576

\bibitem[{Kauffmann} \& {Haehnelt}(2002){Kauffmann} and
  {Haehnelt}]{Kauffmann02}
{Kauffmann}, G., \& {Haehnelt}, M.~G. 2002, \mnras, 332, 529

\bibitem[{Kitzbichler} \& {White}(2007){Kitzbichler} and {White}]{Manfred06}
{Kitzbichler}, M.~G., \& {White}, S. D.~M. 2007, \mnras, 376, 2

\bibitem[{Kocevski} {et~al.}(2008){Kocevski} {\em et~al.}]{Kocevski08}
{Kocevski}, D.~D., et~al. 2008, arXiv:0809.2091

\bibitem[{Laird} {et~al.}(2009){Laird} {\em et~al.}]{Laird09}
{Laird}, E.~S., et~al. 2009, \apjs, 180, 102

\bibitem[{Lidz} {et~al.}(2006){Lidz}, {Hopkins}, {Cox}, {Hernquist}, and
  {Robertson}]{Lidz06}
{Lidz}, A., {Hopkins}, P.~F., {Cox}, T.~J., {Hernquist}, L., \& {Robertson}, B.
  2006, \apj, 641, 41

\bibitem[{Mainieri} {et~al.}(2002){Mainieri} {\em et~al.}]{Mainieri02}
{Mainieri}, V., et~al. 2002, \aap, 393, 425

\bibitem[{McLure} \& {Dunlop}(2004){McLure} and {Dunlop}]{Mclure04}
{McLure}, R.~J., \& {Dunlop}, J.~S. 2004, \mnras, 352, 1390

\bibitem[{Mihos} \& {Hernquist}(1996){Mihos} and {Hernquist}]{Mihos96}
{Mihos}, J.~C., \& {Hernquist}, L. 1996, \apj, 464, 641

\bibitem[{Miyaji} {et~al.}(2007){Miyaji} {\em et~al.}]{Miyaji07}
{Miyaji}, T., et~al. 2007, \apjs, 172, 396

\bibitem[{Montero-Dorta} {et~al.}(2008){Montero-Dorta} {\em
  et~al.}]{MonteroDorta08}
{Montero-Dorta}, A.~D., et~al. 2008, \mnras, p. 1355

\bibitem[{Morokuma} {et~al.}(2008){Morokuma} {\em et~al.}]{Morokuma08}
{Morokuma}, T., et~al. 2008, \apj, 676, 121

\bibitem[{Mullis} {et~al.}(2004){Mullis} {\em et~al.}]{Mullis04}
{Mullis}, C.~R., et~al. 2004, \apj, 617, 192

\bibitem[{Mushotzky}(2004){Mushotzky}]{Mushotzky04}
{Mushotzky}, R. 2004, In Supermassive Black Holes in the Distant Universe,
  A.~J. {Barger}, ed., volume 308 of {\em Astrophysics and Space Science
  Library\/}, p.~53

\bibitem[{Mushotzky} {et~al.}(2000){Mushotzky}, {Cowie}, {Barger}, and
  {Arnaud}]{Mushotzky00}
{Mushotzky}, R.~F., {Cowie}, L.~L., {Barger}, A.~J., \& {Arnaud}, K.~A. 2000,
  \nat, 404, 459

\bibitem[{Myers} {et~al.}(2006){Myers} {\em et~al.}]{Myers06}
{Myers}, A.~D., et~al. 2006, \apj, 638, 622

\bibitem[{Myers} {et~al.}(2007){Myers} {\em et~al.}]{Myers07}
{Myers}, A.~D., et~al. 2007, \apj, 658, 85

\bibitem[{Naab} {et~al.}(2007){Naab} {\em et~al.}]{Naab07}
{Naab}, T., et~al. 2007, \apj, 658, 710

\bibitem[{Nandra} {et~al.}(2007){Nandra} {\em et~al.}]{Nandra07}
{Nandra}, K., et~al. 2007, \apjl, 660, L11

\bibitem[{Peebles}(1980){Peebles}]{Peebles80}
{Peebles}, P.~J.~E. 1980.
\newblock {The Large-Scale Structure of the Universe}, Princeton, N.J.,
  Princeton Univ. Press

\bibitem[{Pierce} {et~al.}(2007){Pierce} {\em et~al.}]{Pierce07}
{Pierce}, C.~M., et~al. 2007, \apjl, 660, L19

\bibitem[{Plionis} {et~al.}(2008){Plionis} {\em et~al.}]{Plionis08}
{Plionis}, M., et~al. 2008, \apjl, 674, L5

\bibitem[{Porciani} \& {Norberg}(2006){Porciani} and {Norberg}]{Porciani06}
{Porciani}, C., \& {Norberg}, P. 2006, \mnras, 371, 1824

\bibitem[{Porciani}, {Magliocchetti}, \& {Norberg}(2004){Porciani},
  {Magliocchetti}, and {Norberg}]{Porciani04}
{Porciani}, C., {Magliocchetti}, M., \& {Norberg}, P. 2004, \mnras, 355, 1010

\bibitem[{Richstone} {et~al.}(1998){Richstone} {\em et~al.}]{Richstone98}
{Richstone}, D., et~al. 1998, \nat, 395, A14

\bibitem[{Rovilos} \& {Georgantopoulos}(2007){Rovilos} and
  {Georgantopoulos}]{Rovilos07}
{Rovilos}, E., \& {Georgantopoulos}, I. 2007, \aap, 475, 115

\bibitem[{Salim} {et~al.}(2007){Salim} {\em et~al.}]{Salim07}
{Salim}, S., et~al. 2007, \apjs, 173, 267

\bibitem[{Schawinski} {et~al.}(2007){Schawinski} {\em et~al.}]{Schawinski07}
{Schawinski}, K., et~al. 2007, \mnras, 382, 1415

\bibitem[{Seljak}(2002){Seljak}]{Seljak02}
{Seljak}, U. 2002, \mnras, 334, 797

\bibitem[{Shen} {et~al.}(2007){Shen} {\em et~al.}]{Shen07}
{Shen}, Y., et~al. 2007, \apjl, 654, L115

\bibitem[{Shen} {et~al.}(2008){Shen} {\em et~al.}]{Shen08}
{Shen}, Y., et~al. 2008, arXiv:0810.4144

\bibitem[{Silverman} {et~al.}(2008){Silverman} {\em et~al.}]{Silverman08}
{Silverman}, J., et~al. 2008, arXiv:0810.3653

\bibitem[{Smith} {et~al.}(2003){Smith} {\em et~al.}]{Smith03}
{Smith}, R.~E., et~al. 2003, \mnras, 341, 1311

\bibitem[{Springel}, {Di Matteo}, \& {Hernquist}(2005a){Springel}, {Di Matteo},
  and {Hernquist}]{Springel05a}
{Springel}, V., {Di Matteo}, T., \& {Hernquist}, L. 2005a, \apjl, 620, L79

\bibitem[{Springel}, {Di Matteo}, \& {Hernquist}(2005b){Springel}, {Di Matteo},
  and {Hernquist}]{Springel05c}
{Springel}, V., {Di Matteo}, T., \& {Hernquist}, L. 2005b, \mnras, 361, 776

\bibitem[{Thacker} {et~al.}(2008){Thacker}, {Scannapieco}, {Couchman}, and
  {Richardson}]{Thacker08}
{Thacker}, R.~J., {Scannapieco}, E., {Couchman}, H.~M.~P., \& {Richardson}, M.
  2008, arXiv:0811.2014

\bibitem[{Weiner} {et~al.}(2009){Weiner} {\em et~al.}]{Weiner09}
{Weiner}, B.~J., et~al. 2009, \apj, 692, 187

\bibitem[{Willmer} {et~al.}(2006){Willmer} {\em et~al.}]{Willmer06}
{Willmer}, C., et~al. 2006, \apj, 647, 853

\bibitem[{Yang} {et~al.}(2005){Yang}, {Mo}, {van den Bosch}, and
  {Jing}]{Yang05a}
{Yang}, X., {Mo}, H.~J., {van den Bosch}, F.~C., \& {Jing}, Y.~P. 2005, \mnras,
  357, 608

\bibitem[{Yang} {et~al.}(2003){Yang} {\em et~al.}]{Yang03}
{Yang}, Y., et~al. 2003, \apjl, 585, L85

\bibitem[{Yang} {et~al.}(2006){Yang}, {Mushotzky}, {Barger}, and
  {Cowie}]{Yang06}
{Yang}, Y., {Mushotzky}, R.~F., {Barger}, A.~J., \& {Cowie}, L.~L. 2006, \apj,
  645, 68

\bibitem[{Zehavi} {et~al.}(2005){Zehavi} {\em et~al.}]{Zehavi05}
{Zehavi}, I., et~al. 2005, \apj, 630, 1

\bibitem[{Zheng}, {Coil}, \& {Zehavi}(2007){Zheng}, {Coil}, and
  {Zehavi}]{Zheng07}
{Zheng}, Z., {Coil}, A.~L., \& {Zehavi}, I. 2007, \apj, 667, 760

\end{thebibliography}
\end{document}